\documentclass{jfm}

\usepackage{graphicx}
\usepackage{newtxtext}
\usepackage{newtxmath}
\usepackage{natbib}
\usepackage{hyperref}
\hypersetup{
    colorlinks = true,
    urlcolor   = blue,
    citecolor  = black,
}

\newcommand{\RomanNumeralCaps}[1]
\linenumbers

\title{Self-similar Features in Secondary Breakup of a Droplet and Ligament Mediated Fragmentation under Extreme Conditions}
\author{Saini Jatin Rao\aff{1}
 \and Saptarshi Basu\aff{1}
  \corresp{\email{sbasu@iisc.ac.in}}}
\affiliation{\aff{1}Department of Mechanical Engineering, Indian Institute of Science, Bangalore-560012, India}

\begin{document}
\maketitle

\begin{abstract}
Droplet formation is relevant in many applications spanning natural and artificial settings. A physical understanding of aerobreakup or air-assisted secondary atomization and predicting size distributions in these applications is non-trivial. We show that extreme airflow speeds induce catastrophic breakup, which, although chaotic and seemingly obscure, is not hopelessly unstructured. In the present study, through shockwave-induced breakups, we investigate the associated intermediate processes at smaller spatiotemporal scales at very high Weber numbers ($We \sim 10^3-10^4$). We \textcolor{black}{show} that microscale protrusions decorate the disintegrating droplet interface and eventually fragment, resulting in the generation of daughter droplets. We discover these undulations to follow breakup patterns (sub-secondary breakup) that resemble a scaled-down version of secondary atomization. The consistent topology across a vast range of scales $(10^{-6}m-10^{-2}m)$ suggests a self-similar mechanism bridged by local Weber number. The normalized size distribution of the resultant droplets exhibits universality and $We$ invariance \textcolor{black}{within the investigated range of extreme conditions}, including transient statistics for subsequent time periods. This conforms to a modified gamma distribution characterized by ligament shape factors, which tend toward the limiting behavior associated with the maximum corrugations physically possible. Scaling laws based on the $We$ are derived for the averaged diameter as $\sim {We}^{-1/3}$, using a high-energy aerodynamic breakup mechanism and subsequently used to derive a time-integrated distribution. These observations reinforce the idea of a self-similar mechanism for the catastrophic droplet breakups, encompassing multiscale deformation cascades, self-similar sub-secondary breakups, maximally corrugated ligaments, and universal droplet size distributions.

\end{abstract}

\begin{keywords}
Drops, Breakup/coalescence, Aerosols/atomization
% Authors should not enter keywords on the manuscript, as these must be chosen by the author during the online submission process and will then be added during the typesetting process (see \href{https://www.cambridge.org/core/journals/journal-of-fluid-mechanics/information/list-of-keywords}{Keyword PDF} for the full list).  Other classifications will be added at the same time.
\end{keywords}

% {\bf MSC Codes }  {\it(Optional)} Please enter your MSC Codes here

\section{Introduction}
\label{sec:intro}
\begin{figure}
  \centerline{\includegraphics[width=1\linewidth]{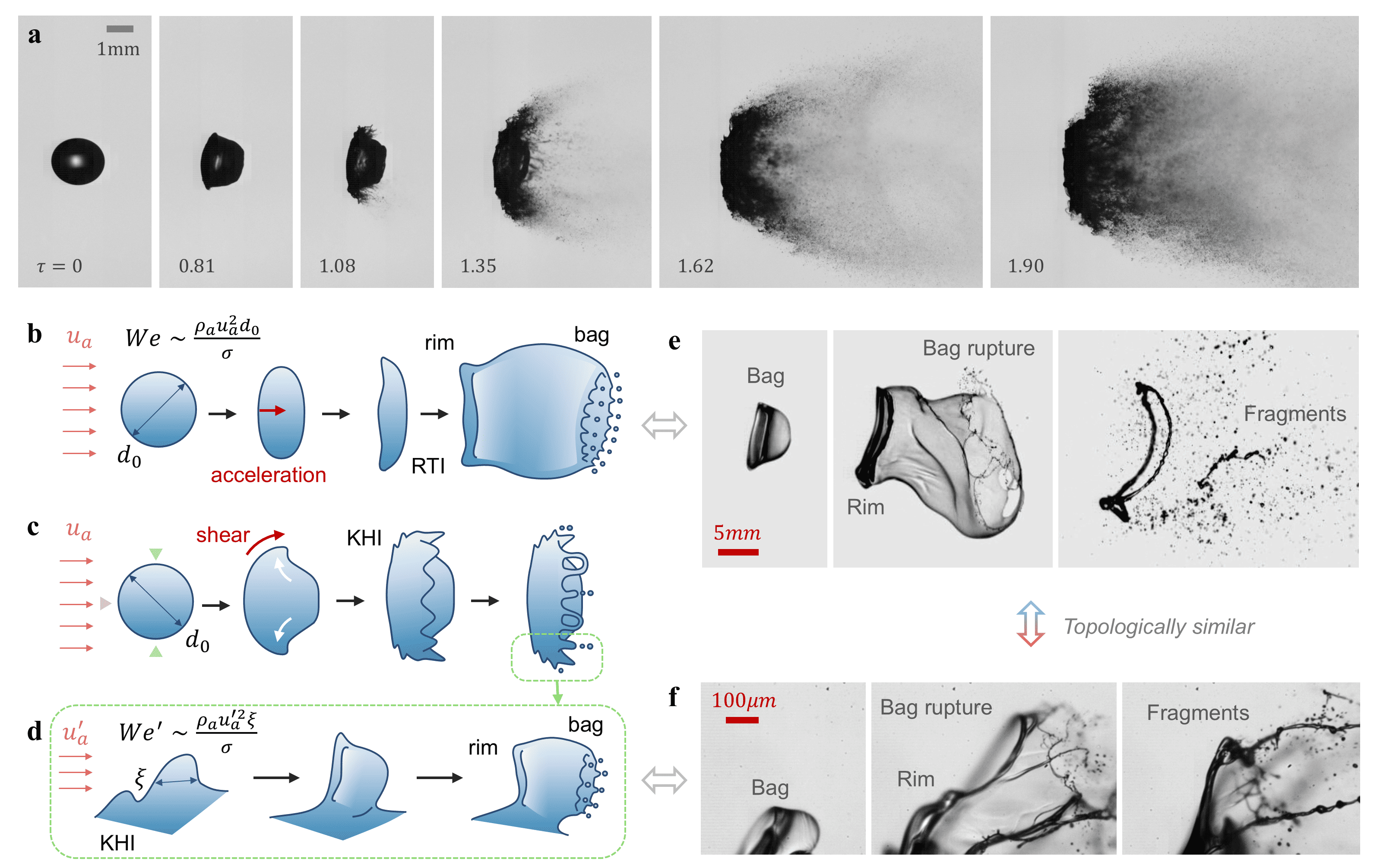}}
  \caption{(a) Shadow images of a shock induced breakup of a droplet at $We\approx2000$, depicting an SIE mode. Schematic illustrating (b) Rayleigh-Taylor Piercing (RTP) (c) Shear-Induced Entrainment (SIE) (d) An undulation fragmentation through sub-secondary breakup process (e) Shadow images of bag breakup of a droplet at $We\approx15$, depicting an RTP mode. This figure adapted from \citet{chandraAerodynamicBagBreakup2024} (f) Shadow images of a sub-secondary breakup process of a droplet, showing the bag mode at much smaller length scales. }
\label{fig:self-similar}
\end{figure}

A liquid drop undergoes rapid destabilization and fragmentation into many tiny daughter droplets when subjected to aerodynamic forcing from high-speed gas flows. This process of generating droplets or secondary atomization in high-speed flow conditions \citep{raoSecondaryAtomizationDroplets2025c} is relevant in many practical scenarios, ranging from energy, aerospace, and aviation applications, where fuel is injected into combustion chambers, emulating extreme scenarios, including shocks and hypersonic flows. The sizing and dispersion characteristics of the fuel control the combustion and emission characteristics. Such interactions are also observed in the supersonic flight of aerial vehicles and missiles or hypersonic entry of space capsules, where bow shocks interact with atmospheric droplets (rain) \citep{dworzanczykAerobreakupStagnationRegion2025}. Understanding this fragmentation process enables the assessment of impact forces, which facilitates the design of effective shields. Furthermore, in spray coating systems and powder production for pharmaceutical, food, or additive manufacturing applications, intense flow conditions are reproduced at the nozzle, and regulating fluid fragmentation is an essential requirement  \citep{sharmaAdvancesDropletAerobreakup2022}. Even natural processes, ranging from volcanic eruptions \citep{jonesFluidDynamicInduced2019} to sea spray \citep{deikeMassTransferOcean2022}, produce droplets through chaotic processes. Determining droplet size and dispersion is a common necessity across these diverse applications, highlighting the need to establish the physics of fragmentation in extreme regimes and develop predictive frameworks. To begin with, an interactive evolution of disruptive and resistive forces establishes the dynamics of breakup. The Weber number ($We=\rho_a u_a^2d_0/\sigma$) represents the balance between aerodynamic and capillary forces, while the Ohnesorge number ($Oh=\mu_l/\sqrt{\rho_l\sigma d_0}$) signifies the dominance of viscous effects. Here $d_0$ is the initial droplet diameter, $u_a$ is the free stream air (or gas) velocity, $\sigma$ is the surface tension for the fluid pair, $\rho$ is the density, and $\mu$ is the viscosity with subscripts $a$ and $l$ representing air (gas) and liquid, respectively. These dimensionless numbers are usually sufficient to predict various aspects of this multi-faceted problem \citep{guildenbecherSecondaryAtomization2009a, theofanousAerobreakupNewtonianViscoelastic2011a, sharmaAdvancesDropletAerobreakup2022, raoSecondaryAtomizationDroplets2025c}. 
One notable aspect of interest is the intermediate processes of fragmentation, which involve the transformation of the large parent droplet $(\sim10^{-3}m)$ into many tiny daughter fragments $(\sim10^{-6}m)$, spanning a large range of spatiotemporal scales. The disintegrating droplet exhibits identifiable intermediate topologies or shapes, which represent different breakup modes at different $We$. These modes can be delineated on a $We-Oh$ regime map \citep{hsiangNearlimitDropDeformation1992b, raoSecondaryAtomizationDroplets2025c}. The older convention established modes based on geometric features, such as: vibrational, bag, multimode, sheet-thinning, and catastrophic breakup \citep{guildenbecherSecondaryAtomization2009a}. The vibrational mode $(We<11)$ entails large-scale oscillation and occasional splitting of the parent droplet into large chunks. The bag breakup mode $(11<We<35)$ involves flattening of the drop, followed by deformation into a rim and a bag shape growing in the downstream direction. The bag film ruptures, resulting in small droplets, while the breakup of the rim produces larger droplets. The multimode or multi-bag breakup $(35<We<80)$ involves the formation of multiple bags and other secondary structures, such as stamen or a complex core in the middle. The sheet-thinning or stripping mode $(80<We<350)$ entails the shear-induced removal of liquid from the droplet periphery via sheet disintegration, resembling a jellyfish. The catastrophic breakup $(We > 350)$, as the name suggests, involves chaotic features that generate very fine droplets. This classification spanned only a small range of flow conditions, and everything above $We>350$ was bucketed into a vaguely defined ``catastrophic" category. This interpretation stems from the inability to observe the rapid and chaotic motion, i.e., catastrophe, at the interface. The drop breakup in this regime is displayed in figure~\ref{fig:self-similar}a at $We\approx 2000$. Recent advances illustrated the significant role of interfacial instabilities; hence, reclassifying the breakup modes to Rayleigh-Taylor Piercing (RTP) and Shear-Induced Entrainment (SIE) \citep{theofanousAerobreakupNewtonianViscoelastic2011a} as illustrated in figure~\ref{fig:self-similar}b-c. The deforming droplet interface, where the lighter fluid (gas) accelerates into the denser fluid (liquid), is destabilized through a Rayleigh-Taylor instability (RTI). This leads to the RTP mode that involves formation of rim and bags over the deformed droplet as illustrated in figure~\ref{fig:self-similar}b and an experimental realization by \citet{chandraAerodynamicBagBreakup2024} at $We=15$ as shown in figure~\ref{fig:self-similar}e. At higher $We$, the aerodynamic shear on the outer part of the forward-facing segment or equator (marked by green triangles in figure~\ref{fig:self-similar}c, where pink triangle denotes the stagnation point) of the droplet interface is significant, inducing Kelvin-Helmholtz instability (KHI) and SIE mode (see figure~\ref{fig:self-similar}a,c,f). The KHI waves mediate the liquid transport along the periphery towards the equator \citep{sharmaShockInducedAerobreakup2021c}, which then disintegrates and removes the liquid mass via a ``stripping" process (see figure~\ref{fig:self-similar}c). These instabilities modulate the initially smooth interface, leading to the formation of localized perturbations, which we refer to as undulations. The stripping is achieved by further destabilization and breakup of these undulations to form sheets, ligaments and droplets as coarsely illustrated in figure~\ref{fig:self-similar}d,f. This particular process, along with the associated nonlinear evolution of the interface, is the focus of this paper, which we will term sub-secondary breakup processes. 

% The KHI in the nonlinear regime undergoes subsequent destabilization through other unstable mechanisms forming a wave cascade, that we will discuss below.\\

The initial evolution of the shear-induced KHI can be prescribed within the linear regime for small amplitudes. In a two-dimensional system, when the KHI wave crests reach a sufficiently finite height, they intrusively interact with the gas flow. A local wake-like features develop behind them, enabling a drag-induced deformation and a scenario often refered to as wave breaking \citep{hoepffnerSelfSimilarWaveProduced2011a}. The surrounding flow imposition catapults the elongated crest into a sheet or ligament-like structure, which subsequently disintegrates and creates droplets \citep{jeromeVorticesCatapultDroplets2013a}. This late-stage growth marks the nonlinear evolution of the destabilized interfacial system. A three-dimensional evolution of KHI poses a possible transverse (or azimuthal) destabilization of the wave features through other wave mechanisms, including RTI. This cascading behavior of the appearance of another instability, imposed over the predecessor, is also observed in primary atomization \citep{marmottantSprayFormation2004b}, where the growth of a shear-driven perturbation imposes acceleration to a liquid feature, advancing into the gas (higher density into lower). This leads to an azimuthal RTI and the formation of transverse modulations, which are further elongated into ligaments via aerodynamic assistance. These long, slender liquid structures or ligaments are destabilized through capillary instability or Rayleigh-Plateau instability (RPI) and ultimately generate droplets. For the early breakup of the local interfacial features or lobes in primary atomization simulations at higher $We$, \citet{zandianPlanarLiquidJet2017,zandianUnderstandingLiquidjetAtomization2018,zandianLengthscaleCascadeSpread2019} observed three distinct sub-scale atomization processes: lobe stretching, corrugated lobes, and hole and bridge formation. These processes, involving the three-dimensional, nonlinear evolution of the interface into a lobe or a locally deformed interfacial feature, ligament, and droplets, are presented as an atomization cascade. A regime map for the topological evolution and breakup of these local features was deduced based on global flow parameters, primarily using a global Weber number and Reynolds number. Furthermore, the temporal evolution of the representative length scale associated with the disintegrating liquid structures, as presented using the averaged radius of curvature of the interface, showed that the system evolved towards diminishing scales \citep{zandianLengthscaleCascadeSpread2019}. This also indicated towards the formation of ligaments prior to disintegration into daughter droplets. This cascade behavior of entwined interfacial deformation and associated neighboring bulk flow features was explained using an approach based on vortex dynamics \citep{zandianUnderstandingLiquidjetAtomization2018}. \citet{wangSimilarityPrimarySecondary2008a} experimentally investigated the lobe or undulation breakup over the destabilized jet surface to be analogous to secondary atomization, where the undulation can be treated as the segment of a droplet. Based on this analogy, they defined a local effective Weber number $We'$ based on undulation size and local flow conditions, to establish a regime map that accommodates different possible modes. The observed modes were morphologically similar to the isolated droplet breakup modes at lower $We$. A transition from ligament- to membrane-mediated breakup mode was identified at around $We'\sim13$, where the latter is synonymous with a bag breakup. This transition aligns well with the established criterion marking the onset of bag breakup in isolated drops, i.e., $We\sim11$ \citep{guildenbecherSecondaryAtomization2009a}. These observations support a cascade mechanism for atomization with self-similar elements, and in the present exposition, we extend these ideas to even smaller scales involved in the catastrophic breakup of droplets. This self-similar paradigm across scales is illustrated in figure~\ref{fig:self-similar}b-f, where small-scale bags in catastrophic breakup is morphologically consistent with bag breakup of a droplet, with an effective Weber number bridging these events.

These sub-scale processes or events occurring over a scale much smaller than droplet size, associated with atomization of the undulations, end with ligament-mediated mechanisms. These ligaments play a vital role in final droplet production through the RPI \citep{marmottant2004fragmentation,castrejon2012breakup,driessen2013stability,palStatisticsDropsGenerated2024a}. The geometric characteristic of the ligament determines the size distribution of the daughter droplets. A model \citep{marmottant2004fragmentation,villermauxLigamentMediatedSprayFormation2004} establishes this correlation by considering aggregation of possible breakup and coalescence events within the sub-elements of the ligament, where a corrugated ligament is imagined to be constituted of blobs (or protoblobs) matching the local diameter. A correlation was then established between the shape factor derived from the statistics of the protoblob sizes within the ligament, and parameters associated with the final droplet sizes, described by the gamma distribution. This approach was further extended to consider the distributed ligaments with different average diameters and determine the parameters of a compound gamma distribution describing the daughter fragments \citep{villermauxDropFragmentationImpact2011,kooijWhatDeterminesDrop2018}. The factors describing these ligament and droplet statistics will be detailed later. This mechanism was further extended to non-Newtonian liquids, where pronounced corrugation on ligaments appeared through viscoelastic mechanisms \citep{keshavarzLigamentMediatedFragmentation2016}.  A limit based on volumetric and geometric constraints was established for extreme levels of corrugation, and corresponding parameters describing the distributions were deduced \citep{eggersPhysicsLiquidJets2008, keshavarzLigamentMediatedFragmentation2016}. This limit is crucial since the present exposition deals with aerobreakup in extreme conditions, and we expect the corrugations to saturate to their maximum limit.

Many studies have spanned these extreme flow conditions \citep{nichollsAerodynamicShatteringLiquid1969,salauddinDetonationShockinducedBreakup2023}; however, there is limited information on the evolution of small-scale interfacial features in the shock-induced catastrophic breakup of a drop, with only a few studies providing clear visualizations of this phenomenon. Additionally, measuring the sizes of fragmented droplets is challenging, considering their extreme attributes: very small size, concentrated spatial distribution, and very high velocity. Many traditional techniques fail, and hence only a few studies have been able to retrieve these statistics using unconventional means, where capturing information of the droplets dispersed over a volume is necessary for a better sampling \citep{tropeaOpticalParticleCharacterization2011,raoDepthDefocusTechnique2024a}. \citet{hsiangNearlimitDropDeformation1992b, hsiangDropPropertiesSecondary1993} measured the distributions from holograms and illustrated that they follow universality up to a certain extent through a normalized root normal probability distribution \citep{simmonsCorrelationDropSizeDistributions1977a}. However, there was no clear physical basis for such distributions, and deviations were also observed in certain cases. \citet{kamiyaStudyCharacteristicsFragment2022a} characterized the transient nature of the droplets produced in shock-droplet interaction and demonstrated that fragments follow a root normal distribution. \citet{sharmaDepthDefocusTechnique2023a} employed a Depth from Defocus (DFD) technique and presented a lognormal probability distribution fit to the transient quantitative measurements. A lognormal distribution originates from the maximum-likelihood or maximum-entropy formalism, drawing inspiration from turbulent cascades and statistical thermodynamics \citep{sellensSimplifiedPredictionDroplet1986, villermauxFragmentation2007, rimbertCrossoverRayleighTaylorInstability2011a, kuoMaximumEntropyFormalism2022a}. Some of these descriptions involve discrete events in which a droplet divides randomly into smaller droplets, followed by subsequent splits, resulting in a breakup cascade. However, catastrophic atomization entails stripping, which continuously generates daughter droplets, engaging in a more continuous description.

In this paper, we assess previously unexplored small-scale breakup events associated with undulations or sub-secondary atomization processes through high-resolution visualization of shock-induced atomization, along with the statistics of the daughter fragments. These sub-scale processes illustrate a subsequent appearance of instabilities superimposed over the predecessor, forming a deformation cascade that eventually terminates with corrugated ligaments that fragment into daughter droplets. The breakup of undulations collectively constructs the stripping process. In the extremity posed by shock-induced atomization, the ligament roughness and the associated transient daughter droplet size distribution are observed to approach the extreme corrugation limit. This enables us to move towards a self-similar model for droplet disintegration in the catastrophic breakup regime. The self-similarity encompasses the following major aspects: (a) Self-similar breakup topology at multiple scales \textcolor{black}{hypothesized to be} bridged by an effective Weber number or the sub-secondary breakup mechanism (b) Corrugated ligaments with roughness saturated to an extreme limit (c) Self-similar size distribution, universal with respect to aerodynamic strength $(We)$ and time instance.

Alongside the discovery of the self-similar aspects in such extreme breakup regimes, scaling laws relying on $We$ are deduced for droplet statistics, including average diameters and number of fragments, based on the catastrophic interpretation of the disintegration. It is expected that this self-similar hypothesis and deduced scaling laws provide a powerful tool that can be generalized to other processes involving chaotic disintegration of a liquid mass, including primary atomization processes, ocean spray with high wind speeds, and many more.

\section{Experimental Setup and Methods}
\label{sec:exp}
\begin{figure}
  \centerline{\includegraphics[width=1\linewidth]{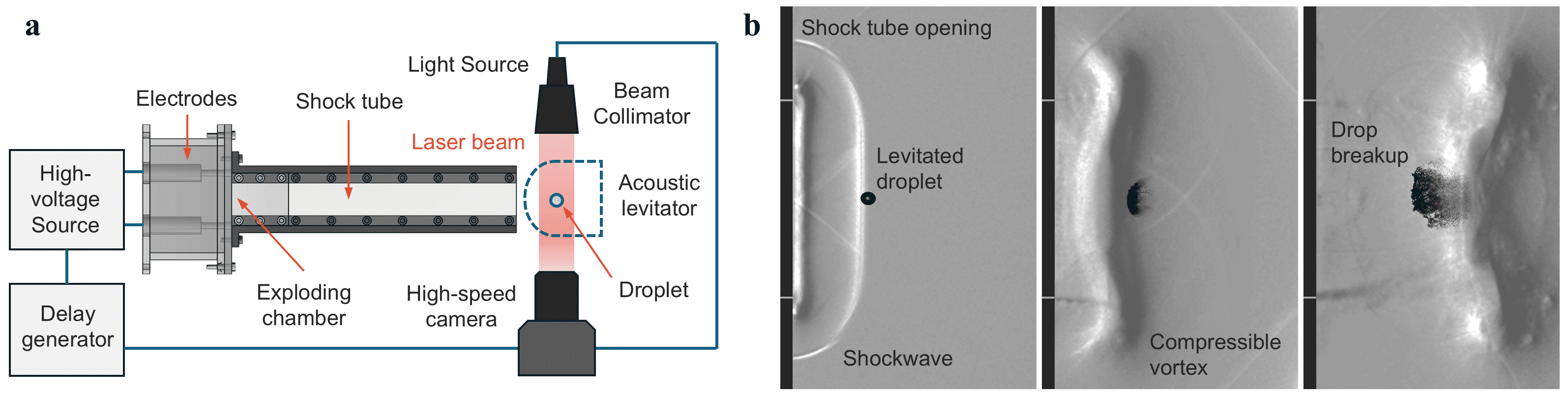}}
  \caption{(a) Experimental setup illustrating a wire-explosion based shock tube, actuated by a pulsed high voltage source. An acoustic levitator is used to place the droplet at the opening, and a shadowgraphy system is deployed for imaging. (b) A droplet interacting with the high-speed flow emerging from the shock tube, illustrating the shock wave and a starting jet.}
\label{fig:setup}
\end{figure}

 An exploding-wire-based shock tube setup was used for the shock-droplet interaction experiments, as illustrated in figure~\ref{fig:setup}a. A high-voltage source (2kJ pulse power, Zeonics Systech India, Z/46/12) is used to discharge electrical energy through a thin wire (35 SWG, bare copper wire) for a very short time duration ($\sim O(1\mu s )$), causing an explosion accompanied by a blast wave. This is transformed into a planar shock wave by deploying a rectangular shock tube cavity ($20mm \times 40mm \times 320mm$). Further details for the same can be found in the previous works \citep{sharmaShockInducedAerobreakup2021c, chandraShockinducedAerobreakupPolymeric2023a, sharmaShockinducedAtomisationLiquid2023a}.  A detailed characterization of the high-speed flow field produced by this device was investigated by \citet{raoBlastWaveInduced2025a}, and the associated shock wave and a starting jet are presented in figure~\ref{fig:setup}b. This system generates a decaying flow, and the necessary details are presented in Appendix~\ref{apx:airflow}.
 An isolated DI water droplet of size $d_0 \approx 2mm$ is placed in the shock tube opening using an acoustic levitator. A digital delay generator (BNC 575) simultaneously triggers the imaging and shock tube system. The shock Mach numbers ($Ma_s$) are controlled by adjusting the charging voltage ($V_c$). Experiments are carried out at $Ma_s=1.27,1.39\ \&\ 1.56$, with corresponding Weber numbers $We\approx900,2000\ \&\ 4000$ \citep{sharmaDepthDefocusTechnique2023a}.

 In the current exposition on the visualization of sub-secondary breakup processes or a closer look at the disintegrating interface, shadow imaging is achieved using a Cavitar Cavilux smart UHS pulsed laser, connected to a beam expander and a diffuser plate for uniform background illumination. The images were captured at $75 kHz$ ($93 kHz$ for some cases) using a Photron SA5 camera. A $6.5\times$ Navitar zoom lens coupled with a $1.5\times$ attachment and $2\times$ objective ensured a resolution of $2.38\mu m$/pixel. 
 
 A modified version of this shadowgraphy-based imaging system was deployed to determine daughter droplet sizes using the Depth from Defocus (DFD) approach, further details of which can be found in the previous study \citep{sharmaDepthDefocusTechnique2023a}. A beam splitter and two high-speed Photron SA5 cameras were used to acquire two simultaneous images with different levels of blurring, i.e., the planes of focus of both cameras were deliberately misaligned, resulting in a prescribed inter-plane spacing. An optical configuration with higher magnification and resolution of $1.4 \mu m/$pixel, achieved using a Navitar $12\times$ Ultra zoom lens coupled with a $4\times$ microscopic objective (Olympus Plan $4\times$), is used to derive reliable estimates of droplet sizes. A calibration procedure was performed by translating target dots of known dimensions along the optical axis at predetermined depth intervals relative to the focal plane of both imaging systems. This process yielded pairs of images for each target in the two cameras, exhibiting variable degrees of image blur due to the differing axial positions of each target relative to the respective camera’s plane of focus. These images were normalized so that gray-level intensities are rescaled to [0, 1], with 0 representing background and 1 representing shadows (droplets). Based on a reference threshold intensity (0.6 in the present study), the effective size of the blurred particle image was estimated at each position, yielding a calibration function for each camera, with threshold-based diameter as a function of depth. Based on these, a triangulation-like approach was implemented to determine the actual size and depth position of the particle \citep{zhouSprayDropMeasurements2020, sharmaDepthDefocusTechnique2023a}. This enabled accurate characterization of particle sizes within a defined measurement volume, thereby improving sampling rates. Additionally, the measurement volume is dependent on the droplet sizes being examined. For reference, the critical depth of measurement is directly proportional to the drop size, and multiplying it by the area of the imaging region yields the measurement volume $V_{\rm meas}$. To ensure accurate results, this information was used to make appropriate bias corrections. By considering the number of droplets per unit measurement volume, unbiased size distributions and probability density functions (PDFs) were determined. More details on the implementation algorithm can be found in \citet{sharmaDepthDefocusTechnique2023a}. \textcolor{black}{The uncertainty in the size measurements depends on the droplet size as well as on its depth position relative to the plane of focus, i.e., the level of defocus. The present optical configuration displayed an uncertainty in sizing of less than $\pm 10\%$ \citep{sharmaDepthDefocusTechnique2023a} and within these constraints, a minimum droplet size of $10\mu m$ was measured. Limitations associated with such measurement techniques are further discussed and generalized by \citet{raoDepthDefocusTechnique2024a}.} The sizing statistics obtained are subsequently assessed here for insights into the fragmentation process.

% \section{Results and Discussion}
% \label{sec:exp}
 
\section{Global overview}
A global view of the shock droplet interaction at a $We\approx2000$ is depicted in figure~\ref{fig:self-similar}a. The time $(t)$ is clocked from the moment the shockwave interacts with the droplet, and it is normalized with the inertial time scale $t_i = d_0/u_{as}\sqrt{\rho_{as}/\rho_l}$ such that $\tau = t/t_i$ \citep{nichollsAerodynamicShatteringLiquid1969}, where $\rho_{as}$ and $u_{as}$ are the air properties just behind the shock and the same has been used to determine overall Weber numbers $We$. In this regime, we observe the SIE mode of breakup. The shock wave itself has minimal influence on the droplet deformation in the present case \citep{sharmaShockInducedAerobreakup2021c}, while the induced flow behind the shock imposes aerodynamic forces and prominent disruption. The drop starts deforming in the presence of non-uniform shear and pressure on its surface. It deforms to a cup-cake shape with a lip-like feature at the equator due to shear and wake \citep{sharmaShockInducedAerobreakup2021c}, as seen at $\tau\approx0.8$ in figure~\ref{fig:self-similar}a and illustrated in figure~\ref{fig:self-similar}a. Shear on the forward-facing segment induces the liquid mass at the interface to transport towards the periphery at the equator, as indicated by the red-arrow in figure~\ref{fig:self-similar}c. This is accompanied by unstable waves originating from KHI, visible as bumps or undulations over the interface at $\tau\approx1$ in figure~\ref{fig:self-similar}a. \textcolor{black}{\citet{sharmaShockInducedAerobreakup2021c, sharmaShockinducedAtomisationLiquid2023a} measured these waves and compared it with various theoretical models for KHI.}  This liquid is then stripped off and forms a cluster of daughter droplets, dispersed downstream by the high-speed air as observed for $\tau>1$ in figure~\ref{fig:self-similar}a. This process continues; meanwhile, the overall parent drop geometry evolves due to persistent aerodynamic forces and a reduction in effective mass until the atomization event is complete. The mechanism of the stripping action involves shear-induced unstable waves, which form these undulations or lobes that break down through a small-scale mechanism referred to as sub-secondary breakup processes. This process is briefly illustrated in figures~\ref{fig:self-similar}d,f. 

 \begin{figure}
  \centerline{\includegraphics[width=0.7\linewidth]{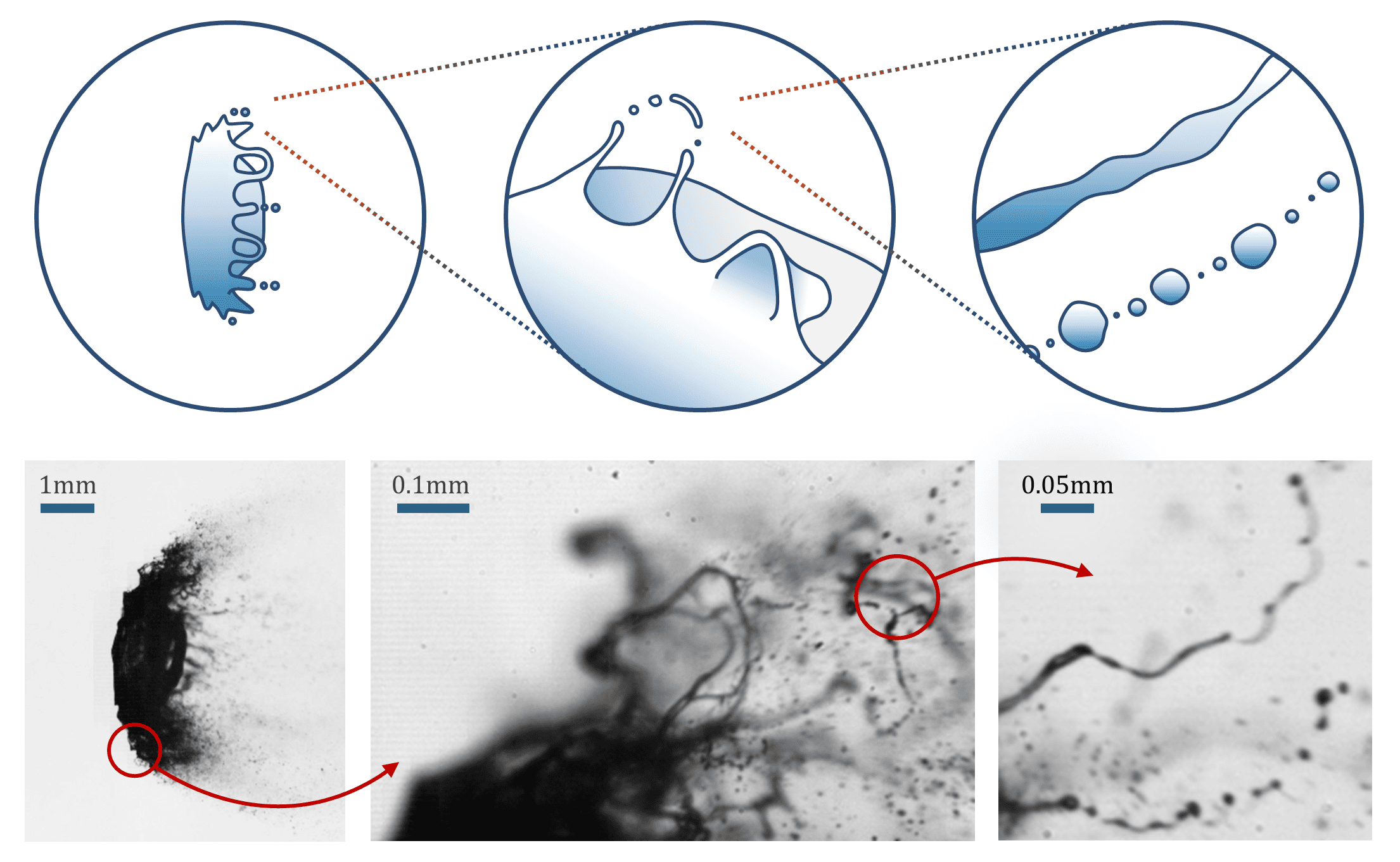}}
  \caption{The deformation cascade happening at a range of spatiotemporal scales, illustrating three prominent stages: global deformations, sub-secondary breakups, and ligament-mediated droplet generation. }
\label{fig:summary}
\end{figure}

A progression of deformations across spatiotemporal scales is illustrated in figure~\ref{fig:summary}. This holistic overview represents the phenomenon as we progressively zoom in on smaller scales. At the droplet level, the deformation spans the overall or global length scale, i.e., the diameter of the droplet, as represented in the first stage of figure~\ref{fig:summary}. As we look closer at the interface, we see that the interface is rough due to KHI. If we zoom in further, the roughness or undulations break down through sub-secondary processes, as represented in the second stage of figure~\ref{fig:summary}, which we will discuss further in the next section (\S~\ref{sec:subsec}). For these sub-scale processes, the deformations span a much smaller length scale, of the order of the undulation size, which is approximately described by the characteristics of KHI. The breakup at this scale is governed by the neighboring or local airflow speeds. If we zoom in further, these processes end with ligaments that finally break down and form the daughter droplets, as represented in the last stage of figure~\ref{fig:summary}. The terminal deformations now span even smaller length scales, of the order of ligament diameters and the resultant daughter droplet sizes. The aerobreakup spans deformations ranging from the global droplet scale to the final daughter droplet sizes, and, as evident from the above discussion, the appearance of these progressively smaller spatiotemporal scales at the interface constructs a deformation cascade. The same can be \textcolor{black}{hypothesized} as progressive superimposition of unstable waves originating from shear, acceleration, or capillary action induced instabilities forming a wave cascade. These sub-scale processes, although random and chaotic, are not hopelessly unstructured. The multiscale deformations as have been vaguely interpreted as a ``catastrophe", will be expanded in the upcoming sections.

% However, to crudely describe, KHI wave crests in the nonlinear regime are destabilized under the influence of local acceleration, and an azimuthal RTI enables periodic ligament formation. These ligaments then generate a fine mist of droplets through RPI, as observed in figure~\ref{fig:self-similar}d.

\section{Sub-secondary breakup processes}
\label{sec:subsec}
\begin{figure}
  \centerline{\includegraphics[width=1\linewidth]{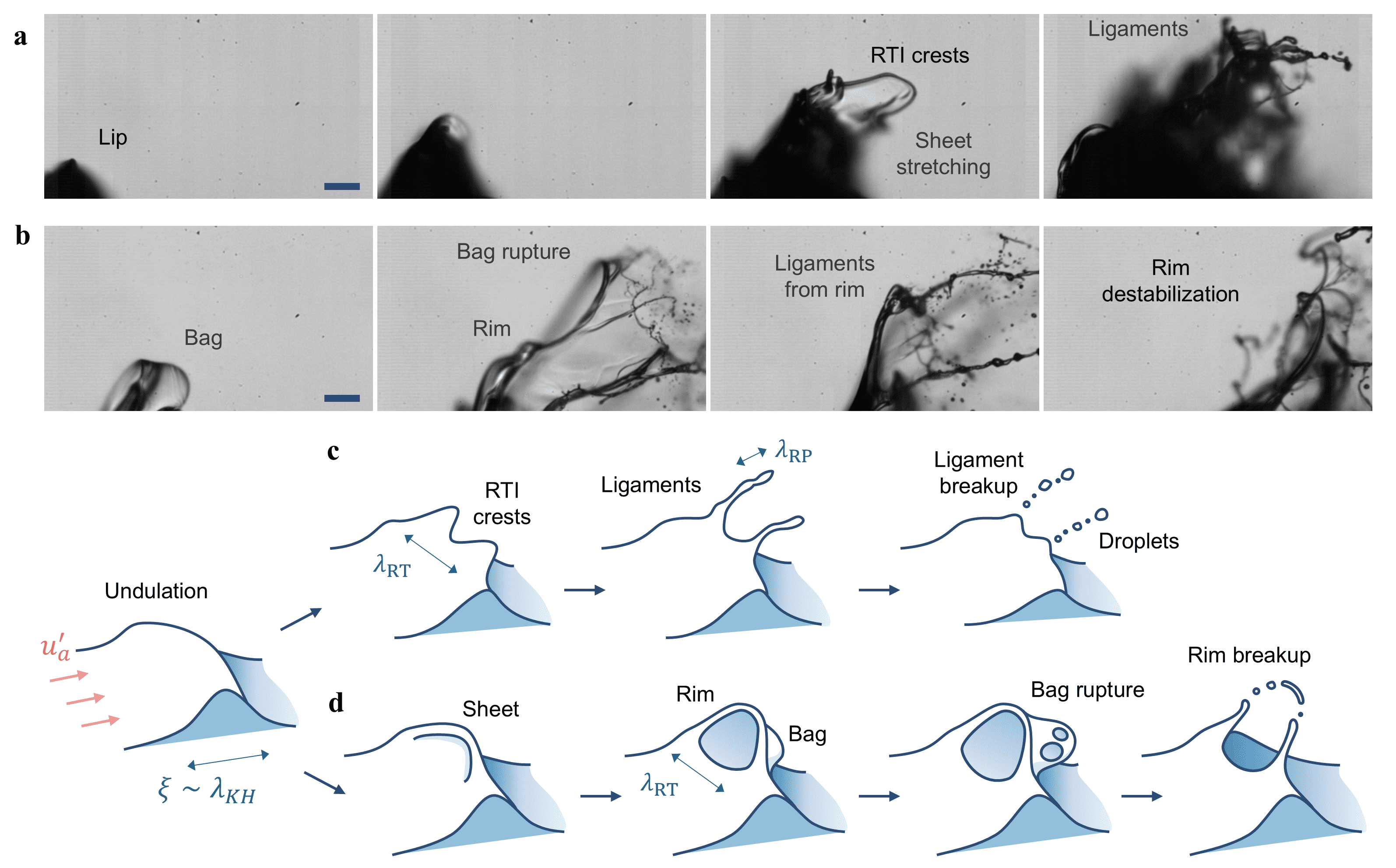}}
  \caption{Evolution of an undulation over droplet surface undergoing sub-secondary breakup processes (a) ligament mode (b) bag mode. The scale bar represents $100\mu m$ and consecutive frames are separated by an interval of $\Delta t=13.33\mu s$. Schematic illustrating sub-secondary breakup processes. The early stage KHI initiates an undulation with characteristic scale $\xi\sim \lambda_{\rm KH}$ with local effective air speed $u_a'$. Beyond a particular finite amplitude, this undergoes destabilization through sub-secondary breakup processes (a) the ligament mode, characterized by transverse acceleration-induced azimuthal RTI, and (b) the bag mode, characterized by longitudinal acceleration-induced azimuthal RTI. Both these modes eventually create terminal ligaments that generate droplets through RPI. $\lambda_{\rm KH}$, $\lambda_{\rm RT}$ and $\lambda_{\rm RP}$ represents wavelength for KHI, RTI and RPI respectively.}
\label{fig:sub-sec}
\end{figure}

\textcolor{black}{A \textbf{primary atomization} process involves the breakup of a bulk or continuous mass of liquid, such as in jets, leading to the formation of large drops, intermediate irregular blobs, or ligaments. Under persistent aerodynamic action, these drops undergo further breakup into smaller daughter droplets, constituting a \textbf{secondary atomization} process. The intermediate processes of interface deformation and mass detachment (fragmentation) during jet breakup bear strong resemblance to secondary atomization \citep{wangSimilarityPrimarySecondary2008a} as discussed in the previous section (\S~\ref{sec:intro}). 
At sufficiently high $We$, however, secondary atomization is not achieved through a single catastrophic breakup event but rather through a multitude of localized stripping events distributed across the evolving interface as isolated, observed, and presented in figure~\ref{fig:sub-sec}a-b. To distinguish these small-scale intermediate elementary events from the overall secondary breakup process, we refer to them as \textbf{sub-secondary atomization}. To formally introduce: Sub-secondary atomization denotes the hierarchical subdivision of a secondary atomization event, wherein the aerodynamically generated interfacial structures (protruded fluid features or undulations) arising on an already deforming droplet undergo a further sequence of deformation and fragmentation that are analogous to secondary atomization itself. Thus, sub-secondary atomization involves the breakup of localized liquid features generated during secondary breakup, rather than the parent droplet as a whole. A complete stripping event in high $We$ secondary atomization is therefore interpreted as the collective spatiotemporal accumulation of numerous sub-secondary breakup events across the liquid interface, as elaborated later in this section. This terminology is introduced here to denote the first hierarchical subdivision within secondary atomization and does not preclude further nested levels of breakup at smaller scales, such as the breakup of corrugated ligaments.}

Since undulations are a precursor to the sub-secondary breakups, it is essential to understand their origins. Aerobreakup of a droplet involves initial deformations at the global scale that set up the stage for further unstable mechanisms to manifest. In the SIE breakup mode, a shear-driven KHI generates approximately periodic perturbations (or undulations) to the interface corresponding to the wave crests with a characteristic wavelength $(\lambda_{KH})$ and growth rates, \textcolor{black}{as characterized experimentally for the same flow configuration by \citet{sharmaShockInducedAerobreakup2021c, sharmaShockinducedAtomisationLiquid2023a} and numerically by \citet{jalaalTransientGrowthDroplet2014d}}. At extreme airflow impositions (high $We$), the deforming parent drop interface imposes a significant acceleration of the lighter fluid into a denser one, inducing RTI on the forward-facing interface, alongside KHI near the equator, \textcolor{black}{as observed by \citet{chandraShockinducedAerobreakupPolymeric2023a} for the same flow configuration.} In such scenarios, both KHI and RTI mechanisms create undulations at the interface. However, the prominence of the mechanisms differs spatially, with RTI occurring near the stagnation point due to significant acceleration, and KHI appearing closer to the equator because of substantial shear in that region \citep{dworzanczykAerobreakupStagnationRegion2025}. Consequently, the undulations prone to stripping observed near the equator are primarily driven by KHI. The wavelength represents the length scale of the undulation and impending deformation/fragmentation, represented by $\lambda_{KH}$ and $\lambda_{RT}$ for KHI and RTI respectively. The wavelength can be deduced using linear stability analysis, enforcing simplistic assumptions but reliable estimates \citep{jalaalTransientGrowthDroplet2014d, sharmaShockInducedAerobreakup2021c}. However, these estimates are valid only in the linear regime for small amplitudes. When the undulation is sufficiently large, the evolution becomes nonlinear, with other unstable waves manifesting over this feature, leading to three-dimensional deformations. \textcolor{black}{\citet{marmottantSprayFormation2004b} established a detailed framework to analyze the subsequent appearance of these instabilities (KHI, RTI and RPI).}

To reemphasize, there is a characteristic spatiotemporal signature associated with these local undulations over the evolving liquid boundary. Further evolution of these undulations, beyond the KHI-induced inception, constitutes the sub-secondary breakup process as shown in figure~\ref{fig:sub-sec}a-b. In figure~\ref{fig:sub-sec}a, the undulation near the equator appears to stretch into sheets which destabilize into elongated ligaments. In figure~\ref{fig:sub-sec}b, the undulation appears to be blown into a bag bound by rims, both of which break down into ligaments and droplets. Both these scenarios illustrate the small-scale breakups, and these processes can be comprehended as the nonlinear evolution of a KHI wave with a fluid pair distinct properties across the interface, including a high velocity jump, drastic density and viscosity contrast, and a prominent surface tension component. The nonlinear progression of KHI in the presence of surface tension is a common feature observed in wave-breaking phenomena across various contexts, including ocean waves and shallow water systems \citep{cokeletBreakingWaves1977}.

When exposed to high-speed gas flows, these undulations continue to emerge, deform, and disintegrate, resembling a portion of an independent fluid blob with an effective airflow condition different from the free stream flow, as illustrated in figure~\ref{fig:sub-sec}c-d. This enables the stripping of liquid mass from the parent droplet body in a recurrent fashion \citep{dorschnerFormationRecurrentShedding2020a}. The fresh surface then undergoes the next deformation cycle, setting up an undulation, and again undergoes a sub-secondary breakup process, as illustrated and discussed later in figure~\ref{fig:stripping}. These sub-secondary breakups can be further categorized into distinct modes: ligament and bag modes. The observed modes are presented in figure~\ref{fig:sub-sec}. These modes and their corresponding underlying mechanisms are detailed below.

% The wavelength should be smaller than the droplet for it to manifest over the surface \citep{sharmaShockInducedAerobreakup2021c, sharmaShockinducedAtomisationLiquid2023a}. Additionally, the growth rate of the wave amplitudes sets up a race among various modes with faster growth rates supporting early appearance. These sets a scale-based compatibility criteria. These wave crests are reminiscent of a local undulation over the interface and will be treated as a sub-scale mass of liquid body (see figure~\ref{fig:self-similar} and \ref{fig:sub-sec}). 

\subsection{Ligament mode}
The inception of undulation is governed through KHI-mediated wave crests, and it grows in the transverse direction relative to the interface and airflow. The transverse accent of the wave crest is associated with the growth rate of KHI and global deformation of the overall droplet. This causes an acceleration-induced body force on this segment of liquid associated with the wave crest. With the heavier fluid pushing rapidly into the lighter fluid, the crest destabilizes, and RTI generates azimuthal wave perturbations over this crest as observed in the third panel of figure~\ref{fig:sub-sec}a and depicted in figure~\ref{fig:sub-sec}c. The continued presence of transverse acceleration further augments these RTI wave crests, which eventually elongate through aerodynamic assistance or drag to form ligaments \citep{marmottantSprayFormation2004b, hoepffnerSelfSimilarWaveProduced2011a, jeromeVorticesCatapultDroplets2013a}. These ligaments elongate further and undergo an RPI-mediated breakup into daughter droplets as illustrated in the last panels of figure~\ref{fig:sub-sec}a,c.

However, there is another possible mechanism. The unsteady, rapid expansion of sheets is shown to be bound by unstable rims that shed droplets through periodic ligaments \citep{wang2021growth}. The local undulation growing transversely can also evolve into a sheet as it stretches under air-assisted drag. The presence of a rim at the leading end of this expanding sheet-like feature is difficult to visualize from shadowgraphy images at this early stage. However, if present, the unsteady sheet expansion induces a coupled RTI-RPI, generating ligament-mediated droplets from the unsteady bounding rim \citep{wangUniversalRimThickness2018a}.

In the current context of catastrophic breakup with large $We$, the ligament mode of sub-secondary breakup is comparatively less frequent than the bag mode, as will be discussed later. For this reason, very few such events were clearly captured in the experiments, making it difficult to draw further deductions. However, the ligament mode is typically observed in events characterized by lower $We$ and has been investigated within the context of primary atomization of jets \citep{marmottantSprayFormation2004b, wangSimilarityPrimarySecondary2008a} and sea sprays \citep{veronSeaSpraySpume2012,troitskayaBagbreakupFragmentationDominant2017}. The common conclusion remains that a dominant acceleration in the transverse direction induces a transverse RTI, superimposed over the predecessor KHI waves.

% \begin{figure} 
%   \centerline{\includegraphics[width=1\linewidth]{Figures/sub-sec-scheme.png}}
%   \caption{Schematic illustrating sub-secondary breakup processes. The early stage KHI initiates an undulation with characteristic scale $\xi\sim \lambda_{\rm KH}$. Beyond a particular finite amplitude, this undergoes destabilization governed by subsequent wave mechanisms, forming a nonlinear wave cascade. The accelerating interface undergoes azimuthal RTI. (a) ligament mode: transverse acceleration RTI (b) bag mode: longitudinal acceleration RTI. This eventually creates terminal ligaments that generate droplets through RPI.}
% \label{fig:sub-sec}
% \end{figure}

\subsection{Bag Mode}
Apart from the transverse growth of the KHI-mediated waves, another possible effect arises from the drag forces over this undulation. This undulation resembles a protrusion over the liquid surface, with a bluff body-like flow around it. This is similar to aerodynamic forcing over an isolated droplet. Aerobreakup of a drop shows axisymmetric dynamics that subsume a bi-fold symmetry, thus also dividing it into two equal halves. This means that a \textcolor{black}{bi-fold} symmetry across a horizontal plane passing through the symmetry axis (aligned with the flow direction) is a subset of the axisymmetric case, and we can assume that the top and bottom halves of the droplet undergo similar dynamics. The undulation resembles one such half \citep{wangSimilarityPrimarySecondary2008a}. The undulation begins to flatten under these effects in the same way an isolated droplet does (see figure~\ref{fig:self-similar}b,e), leading to the formation of a sheet-like feature as illustrated in figure~\ref{fig:sub-sec}d. The drag augments the acceleration imparted to the interface in the longitudinal direction along the direction of the gas flow. It is expected for high-speed gas flow that this imparts RTI over this sheet in the azimuthal direction (direction perpendicular to the flow), which leads to the formation of bags bound by the rims as observed in the first panel of figure~\ref{fig:sub-sec}b and depicted in figure~\ref{fig:sub-sec}d. This mechanism is similar to a bag or multi-bag breakup of a droplet.  The bag grows, and the thin film ruptures, creating very fine droplets as seen in the second panel of figure~\ref{fig:sub-sec}b. The residual ligament fragments are under the combined effect of capillary and aerodynamic forces as seen in the third panel of figure~\ref{fig:sub-sec}b. The rims resemble a loop, which opens up as it breaks, with the remnant portion resembling independent ligaments emerging from the parent droplet body as illustrated in the last panel of figure~\ref{fig:sub-sec}d. This bag mode thus resembles the RTP mode in secondary atomization \citep{jackiwAerodynamicDropletBreakup2021b}. A similar bag-mediated breakup at the bulk liquid surface is reported at the ocean surface, creating sea spray or enabling air-sea exchange processes at extreme wind speeds \citep{veronSeaSpraySpume2012,troitskayaBagbreakupFragmentationDominant2017}. Such bags are also observed in our respiratory passages or trachea during coughing/sneezing events, and the associated film breakup was recently emulated using a cough machine \citep{kantBagmediatedFilmAtomization2023}.   

\subsection{Overview on modes and emergent behavior}
In both of these sub-secondary processes, the RTI mechanism destabilizes the KHI crests in the azimuthal direction and eventually establishes a topology that evolves into ligaments that disintegrate through aerodynamically assisted capillary action. However, the ligament mode originates from a dominant transverse acceleration to the undulations, whereas the bag mode is driven by a dominant acceleration in the longitudinal direction aligned with the flow axis. The bag mode involves the breakup of bag film, generating very small droplets. The sheets are ruptured by the spontaneous appearance of holes bound by unstable rims. These rims shed droplets via ligaments \citep{chandraAerodynamicBagBreakup2024}. As this system is significantly perturbed due to the extreme aerodynamic forcing, multiple holes appear on the bags, and their mergers leave behind a thicker intermediate rim \citep{neelFinesCollisionLiquid2020,agbaglahBreakupThinLiquid2021}. This disintegrates into another class of droplets. For both modes of sub-secondary breakups, ligaments constitute the terminal topological structures preceding the formation of the final daughter fragments, and the primary distinction between modes lies in the specific route leading to final ligament formation.

A higher inertia associated with the airflow is required for this drag to manifest as waves and initiate bag mode. Hence, we observe an abundant occurrence of bag mode in this high $We$ regime of present experiments, with a very rare sight of ligament mode as illustrated in Appendix~\ref{apx:subsec}. This behavior can be explained using effective dimensionless parameters defined at the local undulation level.

\begin{figure} 
  \centerline{\includegraphics[width=1\linewidth]{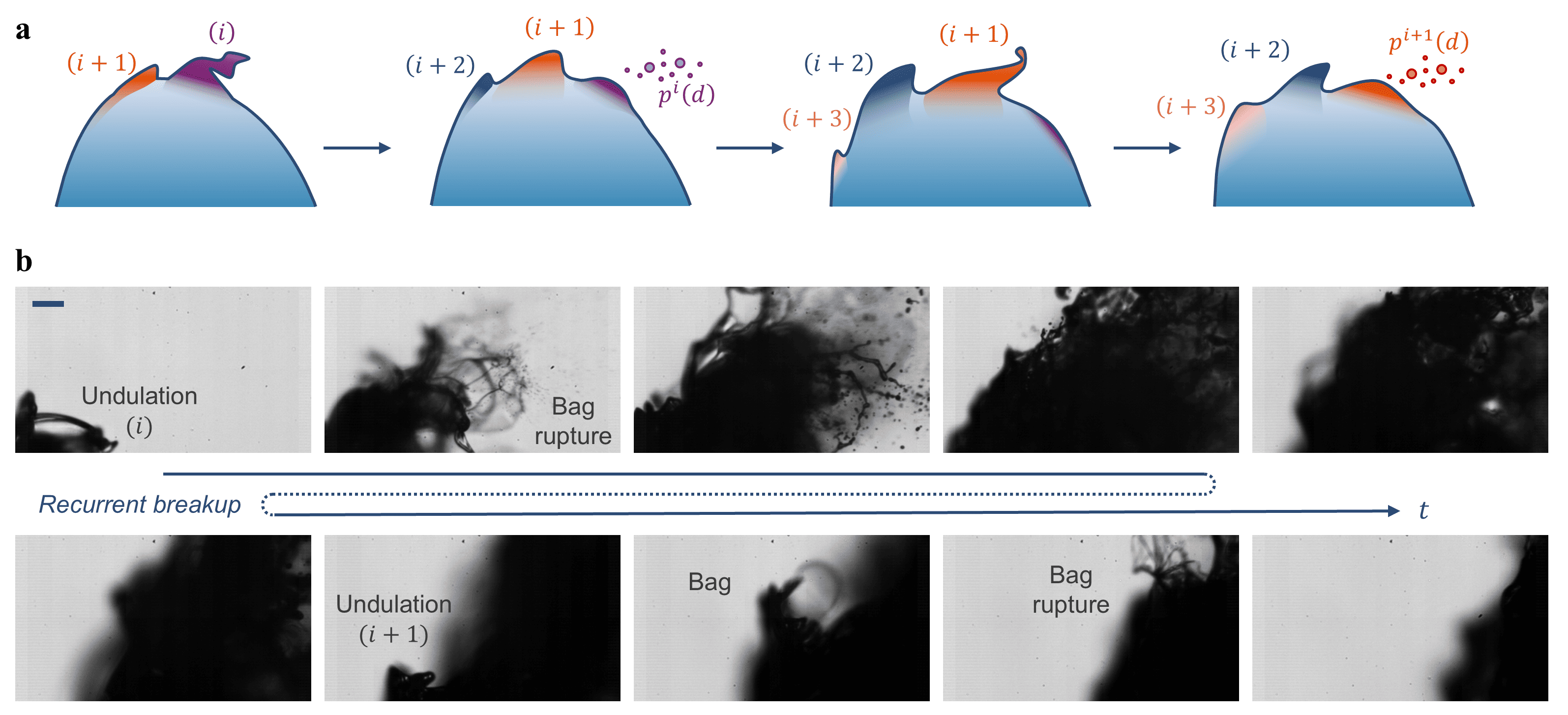}}
  \caption{Striping process in catastrophic SIE as a recurrent sub-secondary breakup mechanism. (a) Schematic depicting the continuous formation of undulations indexed '$i$', and their recurrent breakup generating droplets with distributions $p^i(d)$. (b) Experimental images of the stripping process with recurrent undulation breakup. The scale bar represents $100\mu m$ and consecutive frames are separated by an interval of $\Delta t=13.33\mu s$.}
\label{fig:stripping}
\end{figure}

\begin{figure} 
  \centerline{\includegraphics[width=0.9\linewidth]{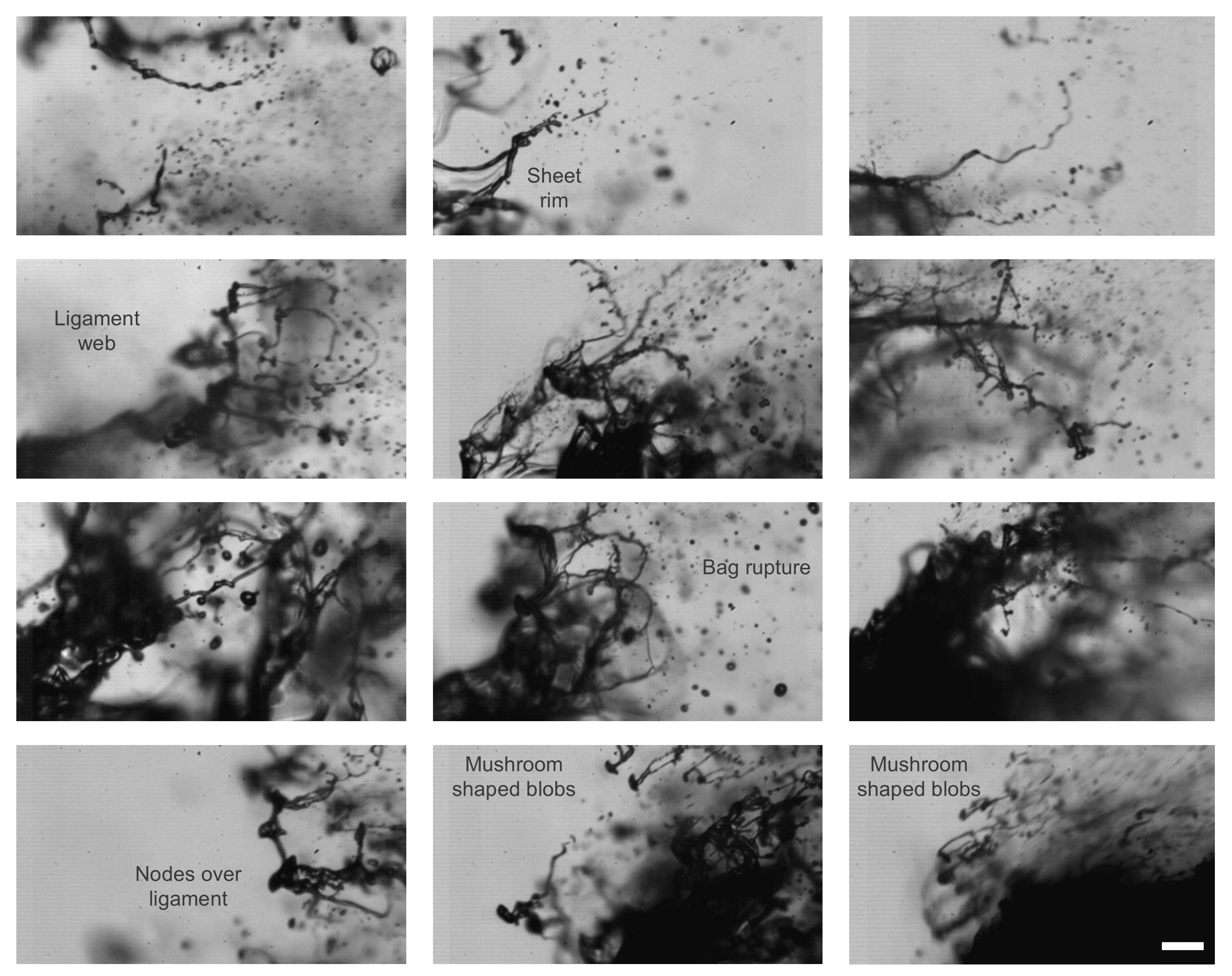}}
  \caption{Corrugated ligaments: Various ligaments observed during the aerobreakup exhibit intricate topologies and significant levels of corrugations. This catalog presents examples of ligaments that remain attached to the parent droplet, as well as those that are free-standing, including mushroom-shaped blobs. Additionally, it includes intra-body ligaments and complex ligament networks that resemble a web structure. The scale bar (bottom right) represents $100\mu m$. The columns from left to right corresponds to $We \approx 900, 2000, 4000$}
\label{fig:ligaments}
\end{figure}

These sub-secondary breakups can be imagined as the elementary fragmentation processes happening at a smaller scale, and the emergent systems depict the stripping of liquid mass from the parent droplet body in a recurrent fashion. This phenomenon is illustrated through figure~\ref{fig:stripping}. The schematic is figure~\ref{fig:stripping}a shows an undulation marked by index '$i$', which breaks through the sub-secondary process, generating droplets with size distribution $p^i(d)$. This exposes a fresh surface on the parent drop, which undergoes further deformation and creation of new undulations. The $i$-th undulation is accompanied by other perturbations indexed $i+1, i+2, i+3 ...$ that undergo further breakups. Recurrent and \textcolor{black}{concurrent} breakup events at smaller scales contribute to overall stripping processes, and when the resultant droplets aggregate, they determine the overall size distribution. \citet{dorschnerFormationRecurrentShedding2020a} observed a similar recurrent breakup pattern, which resembled a ligament shedding event, generating a burst of droplets. \textcolor{black}{Figure~\ref{fig:stripping}b depicts one such instance for recurrent sub-secondary breakups within the experimental observation window. The first undulation is captured partially within the frame of view as tagged by the arbitrary index $(i)$ in the first panel of figure~\ref{fig:stripping}b. This breaks up via the bag mode inferred from the remnants in the second panel. As the system evolves further, a subsequent undulation appears in the seventh panel in temporal sequence (second row, second panel in figure~\ref{fig:stripping}b) tagged index $(i+1)$, which follows a bag mode of sub-secondary breakup as well.}

Therefore, numerous undulations, present over the droplet interface, fragment simultaneously through the aforementioned processes. Such interactions lead to the appearance of extremely corrugated ligaments in complex groups. They sometimes appear as part of a complex web to the parent droplet segments, as illustrated in figure~\ref{fig:ligaments}. Another fascinating observation includes mushroom-shaped short ligaments or blobs resembling the remnant stamen observed in the bag-stamen breakup modes (see last row of figure~\ref{fig:ligaments}). Ligaments that are transversely stretched and aligned perpendicular to the gas flow are subjected to drag, causing RTI-induced breakup, and the remnants of the periodic nodes resemble a mushroom shape. These are similar to nodes and their remnants observed on the rim that appear during the bag breakup of a droplet (including stamen) \citep{jackiwAerodynamicDropletBreakup2021b, chandraAerodynamicBagBreakup2024}.

\textcolor{black}{The parallels of sub-secondary atomization and these modes with established literature for other breakup events were briefly introduced in \S~\ref{sec:intro}. To reemphasize, \citet{zandianPlanarLiquidJet2017,zandianUnderstandingLiquidjetAtomization2018,zandianLengthscaleCascadeSpread2019} observed similar small-scale breakup processes in primary atomization of a jet involving structures like a lobe, which is undulation in the present discussions. The lobes disintegrated and displayed features reported as a hole which corresponds to the bag sheet rupture, and a bridge which is the rim of the disintegrated bag. This corresponds to the bag mode discussed here, and they also reported the formation of a ligament due to lobe elongation, which is in accord with the ligament mode. \citet{marmottantSprayFormation2004b} also emphasized the formation of lobe like structures over the jet surface that elongates under aerodynamic assistance to form ligaments, which subsequently fragment. \citet{wangSimilarityPrimarySecondary2008a} employed x-ray visualization to uncover finer details and established similarities between the undulation breakup over the destabilized jet surface to secondary atomization. Their ligament-mediated and membrane-mediated breakup modes align closely with the ligament and bag mode of sub-secondary breakup discussed here. Similar modes are also reported for ocean spray production in high winds \citep{troitskayaBagbreakupFragmentationDominant2017, veronSeaSpraySpume2012}. These small-scale breakup events that decorate the interface of a breaking liquid surface are extended in the present study to much smaller scales associated with secondary atomization at much larger $We$, and it emphasizes the similarities that span immensely distinct scales. Direct quantitative measurements of these hypothesized consistent breakup morphology, including the subsequent evolution of instability (RTI and RPI) at progressively smaller scales, are necessary to establish an elaborate framework for sub-secondary atomization (see \S~\ref{sec:selfsimilar} for more). The interpretation of KHI and subsequent RTI and RPI in the present discussion on sub-secondary breakups is based on our previous investigations under essentially identical conditions \citep{sharmaShockInducedAerobreakup2021c, sharmaShockinducedAtomisationLiquid2023a, chandraShockinducedAerobreakupPolymeric2023a}, and the extensive body of literature, including \citet{theofanousAerobreakupNewtonianViscoelastic2011a, marmottantSprayFormation2004b, villermauxFragmentationCohesion2020}. However, obtaining such measurements experimentally remains an open challenge because the fragmentation process is inherently three-dimensional and highly transient. Resolving the complete evolution of instability would require simultaneous high spatiotemporal resolution and three-dimensional measurements, potentially through synchronized imaging from multiple views or advanced volumetric techniques. Even these approaches face challenges due to rapid topology changes spanning a large range of scales and limited optical access during fragmentation. Emerging techniques such as time-resolved X-ray imaging may provide additional information and represent a promising direction for future investigations \citep{alisedaXRayFlowVisualization2021}.}

\subsection{Effective local dimensionless parameters}
The deformation cascades originate from the parent drop level up to each KHI-induced undulation and eventually terminate in daughter droplets, as shown in figure~\ref{fig:summary} and \ref{fig:sub-sec}. An effective Weber number $We'$ and Ohnesorge number $Oh'$ can be defined locally for the sub-secondary process based on deformation scale $\xi$ representing the undulation size and local effective gas flow velocity $u_a^{\prime}$ as
\begin{equation} \label{eq:dimensionless}
{We}^\prime\sim\frac{\rho_a u_a^{\prime2} \xi}{\sigma}, \quad \quad \quad Oh'\sim \frac{\mu_l}{\sqrt{\rho_l \sigma \xi}}
\end{equation}

A similar hypothesis bridged primary and secondary atomization events, where undulations over a destabilized liquid jet during breakup were treated as an equivalent droplet segment \citep{wangSimilarityPrimarySecondary2008a}. Based on the local effective Weber number $We'$, the breakup dynamics were predicted for each undulation, drawing inspiration from secondary atomization. The same philosophy is extended here, and is expected to predict the modes associated with sub-secondary breakup process. Other studies for primary atomization \citep{zandianPlanarLiquidJet2017} developed a regime map for potential modes of undulation (lobe) breakup based on global dimensionless parameters. However, a local dimensionless parameter presents a self-similar approach, which potentially enables a hypothesis for a breakup cascade. 

Thus, ${We}^\prime$ predicts the evolution of undulation via the ligament or bag mode discussed earlier. However, determining this ${We}^\prime$ is beyond the experimental capabilities of the present study, since the undulations are much smaller $(\sim10^{-5}m)$ compared to previous studies. \citet{wangSimilarityPrimarySecondary2008a} relied on the droplets generated from the undulation breakup during primary atomization of the jet to determine the total volume associated. Assuming undulation as a hemispherical segment enabled the determination of the characteristic length scale and the free-stream airflow velocity was then used to determine $We'$. A transition from ligament-mediated mode (equivalent to the present ligament mode) and membrane-mediated mode (equivalent to the present bag mode) was observed to be around $We'\sim 13$, and this transition aligns well with the criterion from initialization of bag breakup at $We>13$ for an isolated droplet \citep{guildenbecherCharacterizationDropAerodynamic2017a}. This presents a self-similar framework that predicts a self-similar breakup topology at different scales or stages of the deformation cascade, bridged by an effective Weber number $We\sim We'$. The viscous effect represented by $Oh'$ is determined again based on the local scales and is expected to be higher since $\xi<<d_0$, hence, $Oh'>Oh$. The complex, unsteady evolution of this multiscale two-phase system poses a significant challenge for estimating the effective local airflow velocity using computational, theoretical, or experimental methods \citep{raoSecondaryAtomizationDroplets2025c}. \textcolor{black}{The droplet disintegration involves significant aerodynamic forces, which involve a close-coupled evolution of the two phases. The airflow around the deforming droplet can also be accommodated in a multiscale coupled framework. The large-scale flow around the droplet is similar to the flow past a deforming bluff body and is concerned with the scale representing the transient global geometry of the droplet. Drawing inspiration from solid bluff body flow suggests that the air will accelerate as it moves from the forward stagnation point to the equator, and the outer flow (flow unaffected by the boundary layer close to the surface) will have a magnitude higher than the free stream flow, following separation, causing wake formation in the trail. However, for a deformable interface, aerodynamic drag accelerates the interface, thereby reducing the relative velocity of the liquid relative to air. Furthermore, the extreme shear or drag induced wrinkling of the deformable interface, which further modulates the airflow at scales much smaller than the global geometry, creates small-scale wakes and affects the boundary layer (BL). The resulting undulation thus faces an effective local airflow affected by multiscale coupled two-phase interaction, including BL on both sides of the interface. An estimate of the global geometry on the local airflow velocity and associated instability has been made earlier by considering a quasi-stationary equivalent ellipsoidal shape representing the forward-facing segment of the droplet \citep{dworzanczykAerobreakupStagnationRegion2025, young2025droplet}. The local velocity estimates are of the order of the free-stream velocity. However, given the turbulent nature, the actual flow requires a transient and statistical representation, in which the undulations themselves modify the flow, creating a non-uniform local velocity field. This necessitates estimating the joint statistics of the undulations and the local velocity field, which offers advantages as elaborated in \S~\ref{sec:generalized}.}

\textcolor{black}{However, the order of magnitude analysis of $We'$ and $Oh'$ can be performed based on the relevant scales. At the moment, a crude estimate can be made for the appropriate effective velocity $u_a^\prime$ for $We’$, where $u_a^\prime\lesssim u_a$. Since the undulation originates from a KHI-mediated wave, we have the undulation length scale as $\xi \sim \lambda_{KH}$. Based on the present range of $We$, it can deduced that $\lambda_{KH}/d_0 \sim O(10^{-2}-10^{-1})$ \citep{sharmaShockInducedAerobreakup2021c,raoSecondaryAtomizationDroplets2025c}. Starting with a simpler representation, $u_a'\sim u_a$ can be assumed, following the outer flow velocity scale around the global geometry \cite{dworzanczykAerobreakupStagnationRegion2025,young2025droplet}. The ocean wave literature uses a scale based on shear velocity $u^\ast /u_\infty\sim  \mathcal{O}({10}^{-1})$ \citep{clauser1954turbulent, troitskayaBagbreakupFragmentationDominant2017}, where one can assume $u_a^\prime\sim u^\ast$ giving  $u_a^\prime/u_a\sim\mathcal{O}({10}^{-1})$. Or using a BL approximation where for an unaffected shear layer thickness $\delta$ and a linear flow profile within \cite{sharmaShockInducedAerobreakup2021c}, we have an approximation $u_a/\delta\sim u_a^\prime/\xi$ giving $u_a^\prime/u_a\sim\xi/\delta\sim\mathcal{O}({10}^{-1}-{10}^0)$ depending on the undulation location and transient history. Thus, it is fair to assume that $u_a^\prime/u_a\sim\mathcal{O}({10}^{-1}-{10}^0)$. Substituting this in equation \ref{eq:dimensionless}, we have $We'\sim O(10^{0}-10^{2})$ and $Oh'\sim O(10^{-2})$. Based on these estimates and the regime map presented by \citet{wangSimilarityPrimarySecondary2008a} for undulation breakup and \citet{guildenbecherSecondaryAtomization2009a} for secondary atomization, a bag mode is expected for the sub-secondary breakup processes, consistent with the observed phenomena. Lower values of $u_a'$ due to the turbulent characteristics, however, are expected to be less common in the current high-speed flow conditions, resulting in a less prominent occurrence of the ligament mode of sub-secondary breakup as highlighted in Appendix~\ref{apx:subsec}. The self-similarity is expected to be preserved for a non-uniform local velocity field, since the corresponding $We'$ accommodates this variation and is hypothesized to reproduce the appropriate breakup mode based on the regime it lies in. However, further rigor in quantitative estimates of the velocity field and undulation geometry is necessary to reinforce this hypothesis. The observation of consistent undulation breakup topology or mode for jet and ocean surface breakups within the suggested local $We'$ bounds suggests possible qualitative evidence for the validity of this hypothesis.}

%However, at the moment, crude estimates can be made for the order of magnitude of $We'$ and $Oh'$ based on the relevant scales of the problem. Since the undulation originates from a KHI-mediated wave, we have undulation length scale as $\xi \sim \lambda_{KH}$. Based on the present range of $We$, it can deduced that $\lambda_{KH}/d_0 \sim O(10^{-2}-10^{-1})$ \citep{raoSecondaryAtomizationDroplets2025c}. Assuming $u_a'\sim u_a$ without loss of generality, using equation \ref{eq:dimensionless} we have $We'\sim O(10^{1}-10^{2})$ and $Oh'\sim O(10^{-2})$. Based on the regime map presented by \citet{wangSimilarityPrimarySecondary2008a} for undulation breakup and \citet{guildenbecherSecondaryAtomization2009a} for secondary atomization, a bag mode is expected for the sub-secondary breakup processes, consistent with the observed phenomena. $u_a'$ may also display lower values due to the turbulent characteristics of the system; however, such occurrences are less common in the current high-speed flow conditions, resulting in a less prominent appearance of the ligament mode of sub-secondary breakup as highlighted in Appendix \ref{apx:subsec}.

These undulations are extremely small, topologically complex, and shadow images present only a two-dimensional projection of the same. Furthermore, the time scales of evolution are very small. Hence, it necessitates exceptionally high spatiotemporal resolution to accurately capture the generated droplets for evaluating their equivalent volume or effective undulation scale employing the methodology outlined by  \citet{wangSimilarityPrimarySecondary2008a}. Additionally, assessing the effective local flow velocity is challenging and requires significant advancements in experimental techniques and multiscale multiphase simulation strategies.

% \begin{figure} 
%   \centerline{\includegraphics[width=1\linewidth]{Figures/6.PNG}}
%   \caption{RTI based Ligament breakup mechanism. (a) Schematic depicting the formation of nodes, which remain as “mushroom” blobs at the end. (b) Experimental images.}
% \label{fig6}
% \end{figure}

\section{Ligament-mediated droplet generation}
\begin{figure} 
  \centerline{\includegraphics[width=1\textwidth]{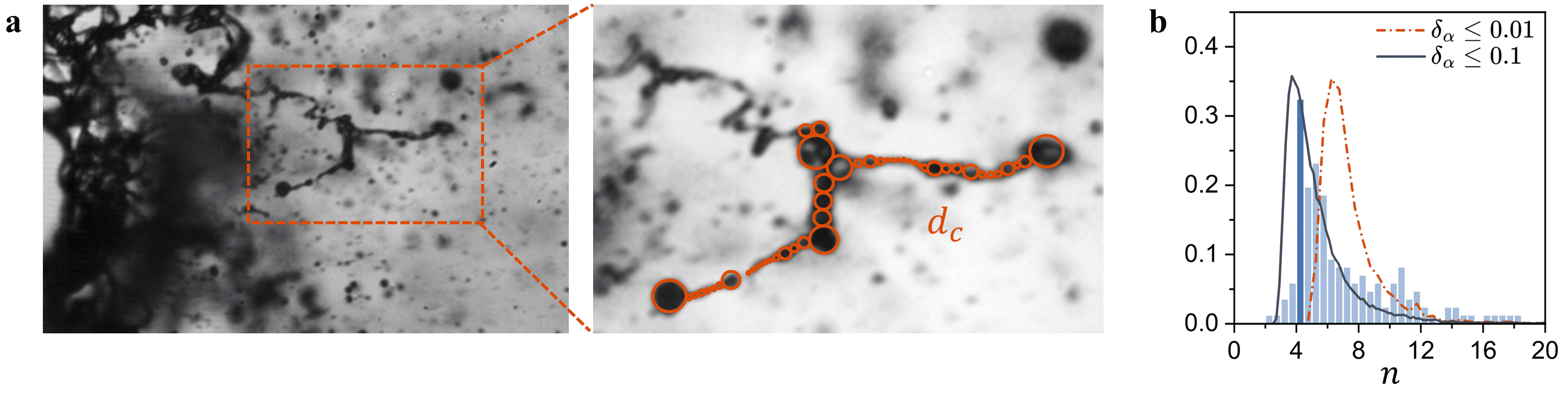}}
  \caption{(a) A typical corrugated ligament decomposed into blobs (marked as circles) matching local ligament diameter $d_c$ (b) PDF illustrating the distribution of ligament corrugation factor $n$ with peak at $n=4$.}
\label{fig:corrugations}
\end{figure}

As emphasized earlier, the sub-secondary breakup processes involve the disintegration of undulation into daughter droplets. The terminal topology just before formation of the droplets entails fine ligaments as illustrated in figure~\ref{fig:sub-sec}, irrespective of the mode of breakup. The corrugated ligaments portray an inertio-capillary system with liquid inertia and surface tension forces dominating the dynamics \citep{marmottant2004fragmentation}. The ligament can be considered as a collection of protoblobs (see figure~\ref{fig:corrugations}) where the evolution ensures two possibilities: either the blobs separate through the RPI mechanism or they coalesce into a bigger blob. Recent studies \citep{marmottant2004fragmentation, keshavarzLigamentMediatedFragmentation2016, palStatisticsDropsGenerated2024a} considered these effects and emphasized the role of corrugations on this delicate balance between breakup and coalescence. These random corrugations arise from the disturbances induced at extreme flow impositions and the coupled interaction between the two phases. A correlation between the ligament shape and the final daughter droplet size distribution has been established in the form of a gamma distribution \citep{villermauxLigamentMediatedSprayFormation2004} or possibly a compound gamma distribution \citep{kooijWhatDeterminesDrop2018}
\begin{equation} \label{eq:compound_gamma}
    P_{m,n}(\bar{d}=d/\langle d\rangle) = \frac{2(mn)^{(m+n)/2} \bar{d}^{(m+n)/2-1}}{\Gamma(m) \Gamma(n)}\mathcal{K}_{m-n}\left(2\sqrt{mn\bar{d}}\right)
\end{equation}
\noindent where $\Gamma$ is the gamma function, $\mathcal{K}_{\mu}$ is the modified Bessel function of the second kind of order $\mu$, and the coefficients $m$ and $n$ represent ligament size distribution and corrugations, respectively.

These coefficients are deduced by decomposing a corrugated ligament into blobs matching local ligament diameters $d_c$ as illustrated in figure~\ref{fig:corrugations}a. Here $d_c$ represents the diameter of the collection of blobs that constitute the corrugated ligament. So, if a ligament is made of $n_b$ number of such blobs, then we have the set of blobs $\left\{d_{c,1},d_{c,2},d_{c,3},\ldots d_{c,n_b}\right\}$  and the averages are then defined as $\langle d_c\rangle= \Sigma_{i=1}^{n_b} d_{c,i}/n_b$ and $\langle d_c^2 \rangle= \Sigma_{i=1}^{n_b} d_{c,i}^2/n_b$. Based on this, $n$ follows the definition $n= \langle d_c\rangle^2/(\langle d_c^2\rangle-\langle d_c\rangle^2)$. This parameter $n$ can be deduced for each ligament before breakup and a distribution (PDF) of the $n$ values for the observed ligaments is presented in figure~\ref{fig:corrugations}b. The coefficient $m=\langle l\rangle^2/(\langle l^2\rangle-\langle l\rangle^2)$ is deduced by taking the mean diameters $l \equiv \langle d_c\rangle$ for individual ligaments, $m$ then describes the distribution of ligament diameters. Given the multiscale nature of the flow, it is expected that the ligaments span a wide range of sizes.

Considering a sample of ligaments spread across space and time with different imposed gas flow $We$, this corrugation factor was determined mostly to be in the range spanning $n=3-12$ as illustrated in figure~\ref{fig:corrugations}b. The values close to $n=4$ were identified as the most frequently observed case for all $We$ considered here. This corresponds to extremely corrugated ligaments, and $n=4$ is a theoretical limit on the coefficient considering various physical constraints \citep{eggersPhysicsLiquidJets2008,keshavarzLigamentMediatedFragmentation2016}. Even the final daughter size distributions, as discussed later in figures~\ref{fig:pdf-map} and \ref{fig:pdf}, follow a PDF with $n=4$. To deduce this limit, let's consider an arbitrary ligament that initially has a smooth surface with radius $R_0$ and is subsequently destabilized, resulting in corrugations. Before the breakup, the corrugated ligament can be considered to be constructed of these protoblobs, with radii $r_i$, where $i$ represents $i$-th blob out of the total $n_b$ blobs, and this is the index over which averaging will be performed. Given that the protoblobs only represent a part of the total ligament volume, we can write the inequality 
\begin{equation}
    \sum_{i=1}^{n_b} \frac{4}{3} \pi r_i^3 \leq \pi R_0^2L 
\end{equation}
Assuming that we represent the initially smooth ligament to be constructed of the same number of protoblobs \citep{keshavarzLigamentMediatedFragmentation2016}, we have $L=n_b\cdot 2R_0$. On averaging, the inequality simplifies to the form $\langle r_i^3 \rangle \leq (3/2) R_0^3$. The corrugation can be represented as perturbations $\xi_i$ such that $r_i=R_0+\xi_i$ and $|\xi_i|\leq R_0$. Furthermore, extending this, a dimensionless parameter $\alpha_i=\xi_i/R_0$ can be defined such that $|\alpha_i|\leq 1$. \textcolor{black}{Substituting this in the inequality, we get 
\begin{equation} \label{eq:corr_volume}
    \langle \alpha_i^3 \rangle + 3\langle \alpha_i^2 \rangle + 3\langle \alpha_i \rangle \leq 1/2
\end{equation}
and recognizing that 
\begin{equation} \label{eq:ncorr}
    n\equiv \frac{1 + 2\langle \alpha_i \rangle + \langle \alpha_i \rangle^2}{\langle \alpha_i^2 \rangle - \langle \alpha_i \rangle^2}
\end{equation}
it gives 
\begin{equation}
    n\geq\  \frac{6}{1- 2\langle \alpha_i^3 \rangle -6\langle \alpha_i \rangle }
\end{equation}
Considering the chaotic nature of the breakup, we can assume the corrugation to be random with $\langle \alpha_i \rangle=0$, which gives $n\equiv 1/\langle \alpha_i^2 \rangle$. Furthermore, enforcing this condition alongside $|\alpha_i|\leq 1$, a variable transformation of the form $\alpha_i=\cos {\theta_i}$ can be considered. Using trigonometric identities, then, we have $\langle \alpha_i^3 \rangle = \langle \cos^3 {\theta_i} \rangle =  (1/4) \langle \cos {3\theta_i} \rangle $. This prescribes $\langle \alpha_i^3 \rangle \in [-1/4,+1/4]$. The limiting value of $n$ is achieved at the most negative value of $\langle \alpha_i^3 \rangle$, giving $n=4$. The PDF of measured $n$ obtained for ligaments sampled from the experimental images is presented in Figure~\ref{fig:corrugations}b. It can be seen that a peak represents that a significant population of corrugated ligaments resides at this extreme limit $n\leq 4$. Yet there is also a small population of ligaments that breach this limit with $n < 4$. Although this deviation is minimal, it is indicative of invalidity of one of the assumptions, especially $\langle \alpha_i \rangle=0$. For random perturbations, the zero mean configuration is statistically the most probable state, as suggested by the central limit theorem, applicable primarily to very long ligaments where the perturbations are symmetric around the base state.}

\textcolor{black}{To assess the effect of the finite length of ligaments, the aspect ratio is defined as 
\begin{equation} \label{eq:aspect_ratio}
    \mathcal{A} = \frac{L}{2R_0} = \frac{2\Sigma r_i}{2\langle r_i \rangle}=n_b
\end{equation}
which shows it also represents the total number of protoblobs. This parameter remains fixed during the evolution of the corrugation immediately preceding breakup, as observed by \citet{keshavarzLigamentMediatedFragmentation2016}. Aspect ratio is essential since ligaments are susceptible to breakup only beyond a threshold $\mathcal{A} > \mathcal{A}_c$ as proposed by various studies \citep{castrejon2012breakup,driessen2013stability, palStatisticsDropsGenerated2024a}. $\mathcal{A}\rightarrow1$ represents droplets and resists further breakup in the absence of external forcing. Returning to the earlier representation of corrugated ligaments through randomly distributed $\alpha_i$ and the assumption $\langle \alpha_i \rangle=0$, even if the random illustration is valid, the zero mean state requires $n_b$ to be very large. Ideally, $\langle \alpha_i \rangle \rightarrow 0$ is expected for infinitely long ligaments, i.e., when $n_b\rightarrow \infty$. Since the ligament lengths are typically finite, limited by the length scale of the undulation that underwent sub-secondary breakup, we expect a small residue $| \langle \alpha_i \rangle | \leq \delta_\alpha$, and it means that the perturbations are not precisely symmetric around the base state.}

\textcolor{black}{The implication of the small residual bias is deduced through a demonstrative Monte Carlo simulation, where a random population of ligaments is generated to assess $n$ statistics. The initial population consists of $10^6$ ligaments, each made up of $n_b=10$ protoblobs, i.e., an aspect ratio of $\mathcal{A}\approx 10$. Each protoblob within a particular ligament is assigned a uniformly distributed random $\alpha_i$ value such that $-1 \leq \alpha_i\leq 1$. The initial ligament population is then filtered imposing inequality presented in Equation~\ref{eq:corr_volume} and residue $| \langle \alpha_i \rangle | \leq \delta_\alpha$. The $n$-PDFs for the filtered feasible ligaments with $\delta_\alpha = 0.01$ and $\delta_\alpha = 0.1$ are displayed alongside experimental data in Fig.~\ref{fig:corrugations}b. A reasonable match is observed even with minimal asymmetry in corrugations, which is surprisingly achieved solely via stochastic processes. Therefore, the asymmetric behavior imposes a spread in $n$ values, including a small population with $n<4$. However, $n=4$ still corresponds to a significant population, making it an appropriate choice for predicting droplet size distributions corresponding to the maximally corrugated state of the system.}

Also, a very low value of $m=3$ is observed, corresponding to the best possible fit to the size distribution to be discussed soon in figure~\ref{fig:pdf} and corresponding fitting details discussed in Appendix~\ref{apx:dist}. This represents a broad ligament size ($l$) distribution, as observed from the experiments as well (see figure~\ref{fig:ligaments}). This illustrates the multiscaled nature of the atomization processes that generate intermediate features over a vast range of scales.

The invariance of size distributions with respect to $We$ number, as discussed in the next section (\S~\ref{sec:dropdist}, see figures~\ref{fig:pdf-map} and \ref{fig:pdf}), supports the saturation of the coefficients $n$ to the theoretical limit. Although the sub-secondary breakup process involves the ligament production through various mechanisms, the ligaments collectively are postulated to exhibit extreme levels of corrugations, primarily due to the highly potent disruptive aerodynamic forces at extreme $We$. The ligaments are expected to scale down to smaller average diameters as $We$ is increased (as observed in the experiments in figure~\ref{fig:ligaments}), but the shape factor is already saturated at $n=4$. To elucidate this hypothesis, the estimated size distribution, predictions from ligament-mediated mechanisms, and deviations are elaborated in the subsequent sections.

\section{Droplet size distribution}
\label{sec:dropdist}
Due to the unsteady imposition arising from this open nozzle-blast wave induced flows \citep{vadlamudiInsightsSpatiotemporalDynamics2024, chandraShockinducedAerobreakupPolymeric2023a, chandraElasticityAffectsShockinduced2024}, the dynamics are expected to be transient. This airflow was recently characterized \citep{raoBlastWaveInduced2025a}, and a power-law decay of airflow was observed, which is discussed in the Appendix~\ref{apx:airflow}. This imposed an unsteady Weber number $We (t)$. Furthermore, the recurrent sub-secondary breakup processes continuously strip liquid away from the parent droplet mass, generating smaller droplets. The impulsive flow also accelerates the remaining parent droplet, causing the effective air velocity to decrease over time, thereby reducing the effective Weber number. Hence, the daughter droplets are produced with evolving characteristics over time, depending on the flow around the residual parent droplet. Therefore, it is necessary to capture the transient nature of fragmentation.

To overcome the limitations of the measurement technique in tackling extremely dense droplet clusters, the measurement window was strategically positioned away from the parent droplet, where the dispersed daughter droplets were sparse enough. This zone was located $30 mm$ downstream from the parent drop for a better visibility of daughter droplets \citep{sharmaDepthDefocusTechnique2023a}. To capture the transient effects, droplet size measurements were performed considering the time stamp of their appearance in the measurement zone. If the first fragments appear in the zone at $t=t_{b0}$, then we define $t^*=t-t_{b0}$, such that $t^*=0$ at the first instant. When the breakup ends, and the last droplets in the zone appear, we set $t^*=t_b^*$, which reflects the total breakup time period. $t^*$ is then normalized using the breakup period $\tau_b=t^*/t_b^*$, such that $\tau_b\in[0,1]$. For the assessment of transient characteristics of the system, this breakup period will be equally divided into sub-periods, and the droplets observed in that temporal window will then be assessed to estimate distributions and average diameters for this particular period (with its center serving as a reference for plotting). So, for instance, if the breakup period is divided into $p$-periods named $T1, T2, ..., Tp$, the first period is given as $T1:\tau_b\in[0,1/p)$ and any arbitrary $i$-th period is then defined as $Ti:\tau_b\in[(i-1)/p,i/p)$, with $i=1,2,...p$.

\begin{figure} 
  \centerline{\includegraphics[width=1\linewidth]{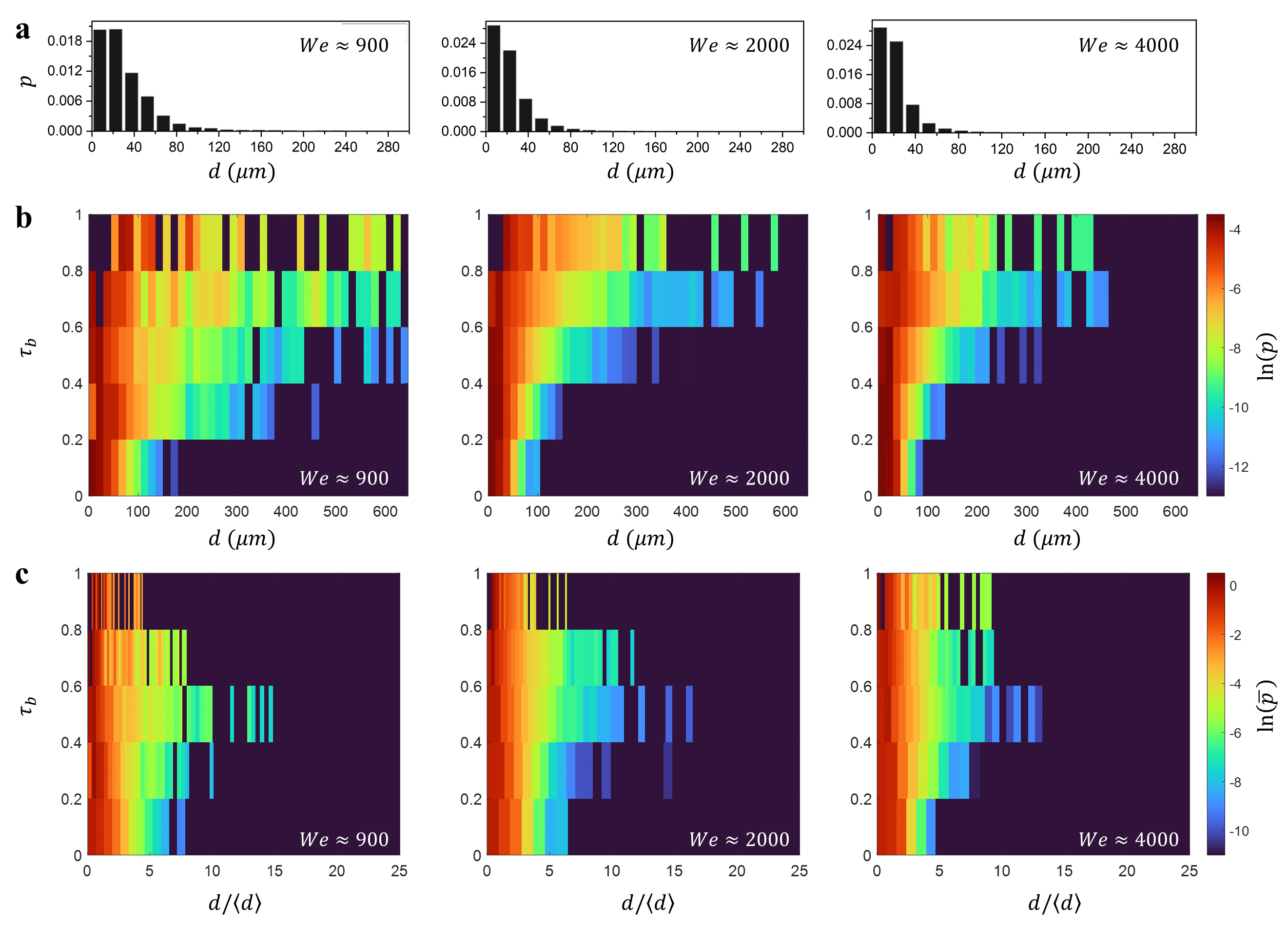}}
  \caption{(a) Overall droplet size distribution PDFs at different $We$. (b) Temporal evolution of PDFs $p(d)$ represented a contour plot, where the breakup event is subdivided into 5 equal periods. (c) Temporal evolution of normalized PDFs $\bar{p}(d/\langle d\rangle)$, using average droplet diameter $\langle d \rangle$, depicting self-similar behavior in time $\tau_b$.}
\label{fig:pdf-map}
\end{figure}

All the sampled droplets at a given $We$ flow, when consolidated for the entire breakup period, gave the overall size distributions for that $We$. The overall distributions or probability density functions (PDFs) $p(d)$ are illustrated in figure~\ref{fig:pdf-map}a. The average droplet size decreases as $We$ is increased. The transient PDFs are represented as a contour plot in figure~\ref{fig:pdf-map}b, where each row (horizontal line) corresponds to the size distribution estimated in that time period (breakup time period $t_b^*$, as discussed earlier, subdivided into equal sub-periods). The contours depict the appearance of larger droplet sizes as time progresses in figure~\ref{fig:pdf-map}b. However, if we normalize the PDFs in each time period with the corresponding average droplet size $\langle d\rangle(\tau_b)$ to obtain the form $(\bar{p}(d/\langle d\rangle))$, we observe the contours to collapse with vertical strips of same colors, which indicates that the different time periods follow the same underlying distribution as illustrated in figure~\ref{fig:pdf-map}c. The same effect is visible across all $We$ discussed in this study. 

\begin{figure} 
  \centerline{\includegraphics[width=0.9\linewidth]{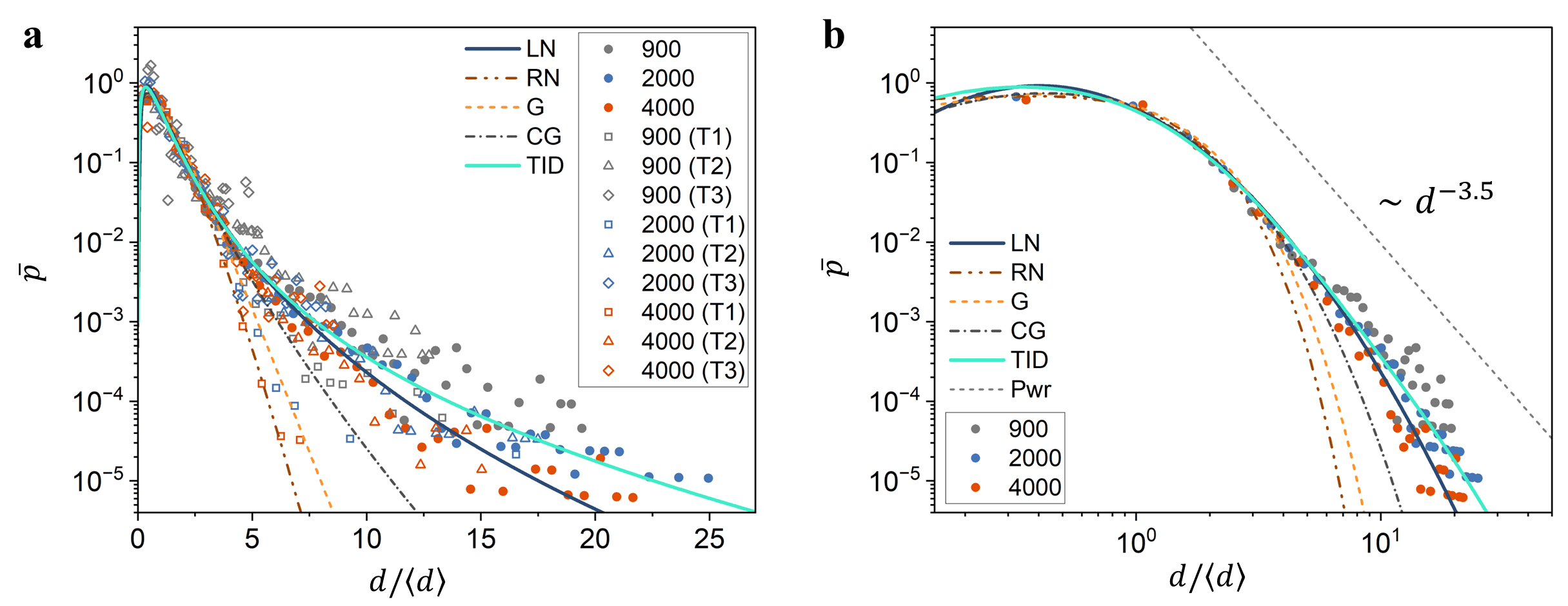}}
  \caption{(a) Normalized PDFs $\bar{p}(d/\langle d\rangle)$ depicting self-similar distributions across $We$ and time (T1-T3 denoting equally divided breakup sub-periods)  (b) The normalized PDFs near the peak in a log-log scale, highlighting a power law tail. Distribution legend: lognormal (LN), root normal (RN), gamma (G), compound gamma (CG), time integrated distribution (TID) and power law (Pwr).} 
\label{fig:pdf}
\end{figure}

For more reliable statistics with a larger sample size in each time period, the overall breakup is split into three temporal parts $T1-T3$, and the corresponding normalized PDFs, along with the global distributions, are presented in figure~\ref{fig:pdf}a. \textcolor{black}{This also includes frequently reported distribution models \citep{lefebvre2017atomization, villermauxFragmentation2007, palStatisticsDropsGenerated2024a} such as lognormal \citep{kuoMaximumEntropyFormalism2022a,tangBagFilmBreakup2023, sharmaDepthDefocusTechnique2023a}, root normal \citep{simmonsCorrelationDropSizeDistributions1977a, hsiangNearlimitDropDeformation1992b, kamiyaStudyCharacteristicsFragment2022a}, and gamma \citep{villermauxLigamentMediatedSprayFormation2004, jackiwPredictionDropletSize2022b} distributions, alongside the compound gamma distribution (equation~\ref{eq:compound_gamma}).} They almost overlap, especially till $d/\langle d\rangle \approx 6$ as highlighted in figure~\ref{fig:pdf}b. The compound gamma distribution aligns well with the prediction of $n=4$, as deduced from the ligament-mediated mechanism discussed earlier. The parameter $m$ is determined from the best-fit to the data. \textcolor{black}{The discussion of the properties of fits and their relative performance is included in the Appendix~\ref{apx:dist}.} The deviation is observed consistently at the tails, which might be due to effects such as (i) the unsteady decaying airflow with lower effective $We$ at later stages results in larger droplets being produced, and (ii) coalescence of daughter droplets. The tail of the PDF represents the population of droplets with large diameters, and despite their lower frequency, they contribute significantly to the volumetric fraction of the fragment. Accordingly, it is imperative to correct this deviation. To gain a better understanding, we require unsteady deformation and instability models. However, an unsteady Weber number-based correction is proposed later within this section. 

Since the normalized PDFs follow a common underlying distribution, predicting the average diameter is ideally sufficient to determine the actual PDFs. From the properties of a gamma distribution, this PDF can be represented in the form

\begin{equation} \label{eq:normalized_pdf}
    p\left(d\right)=\frac{1}{\left\langle d\right\rangle}\bar{p}\left(\frac{d}{\left\langle d\right\rangle}\right)
\end{equation}
 where $\bar{p}\left(d/ \left\langle d\right\rangle;m,n  \right)$ is the universal distribution as displayed in figure~\ref{fig:pdf}a and is independent of $We$ (also see Appendix~\ref{apx:dist}). All that remains is to establish a relationship between $\left\langle d\right\rangle$ and $We$ to reconstruct the distribution. 

\begin{figure} 
  \centerline{\includegraphics[width=1\linewidth]{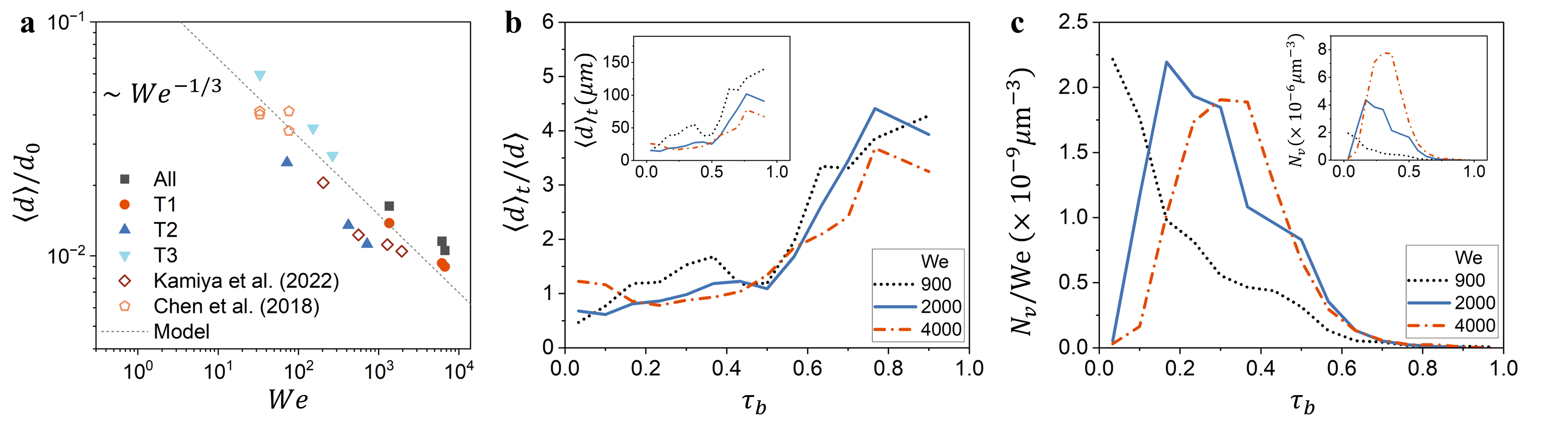}}
  \caption{(a) Variation of normalized overall average diameters $\left\langle d\right\rangle/d_0$ with $We$. Transient evolution of (b) average diameter of daughter droplets $\left\langle d\right\rangle_t/\left\langle d\right\rangle$ with $We$ (c) normalized droplet frequency $N_v/We$. They are normalized using scales based on the catastrophic breakup assumption. The insets denote non-normalized counterparts.}
\label{fig:scaling}
\end{figure}

\citet{sultanovDropletSizeDistribution1990a} proposed a scaling approach where they assumed a chaotic isotropic deformation of the liquid mass to analytically deduce size distributions in explosive breakups associated with a high level of disruptive energy originating from explosions. We extend the same ideas to present a catastrophic aerobreakup scenario. The imposed disruptive energy $E$ induces a deformation rate of $\dot{\gamma}$ to the droplet of initial radius $r_0=d_0/2$, such that 
\begin{equation}
    (1/2)\rho_l r_0^3 (\dot{\gamma} r_0)^2 \sim E
\end{equation}
Assuming chaotic random dispersal \citep{sultanovDropletSizeDistribution1990a} or an energetic isotropic deformation, we assume that the same deformation rate can be extended to the smaller scales where fragmentation actually occurs, and new surface area is generated. This implies we extend this chaotic deformation to individual daughter droplets with characteristic size $r_f = d_f/2$ (radius), such that the associated incremental surface energy scales as
\begin{equation}
    (1/2)\rho_l r_f^3 (\dot{\gamma} r_0)^2 \sim \sigma r_f^2
\end{equation}
Simplifying these equations where $E\sim \rho_a u_a^2r_0^3$ for aerobreakup using high-speed gas flow, we get $(r_f /r_0)\sim {We}^{-1/3}$. Without loss of generality, assuming $\left\langle d\right\rangle\sim d_f$, we have
\begin{equation}
    \left\langle d\right\rangle/d_0\sim {We}^{-1/3}
\end{equation}
\textcolor{black}{This scaling relationship is validated in figure~\ref{fig:scaling}a, where by fitting a form $\left\langle d\right\rangle/d_0 \propto {We}^{\phi}$ we observe $\phi= - 0.346 \pm 0.088$.} Instead of a global $We$, the instantaneous $We$ at the beginning of each time period is considered here for $T1-T3$. The detailed approach to evaluating this is discussed in Appendix~\ref {apx:airflow}. Although the observed breakup is not completely random or isotropic, the scaling hypothesis holds. The averages from available data in a handful of studies that measured such distributions of fragment sizes from shock-induced droplet breakup by \citet{kamiyaStudyCharacteristicsFragment2022a, chenGalinstanLiquidMetal2018} are also included in figure~\ref{fig:scaling}a, which agrees with the present model. Deviations from this are expected at lower $We$, where the breakup is evidently regular in geometry and prominently anisotropic. A similar scaling law for average droplet diameters $(\sim {We}^{-1/3})$ was deduced by \citet{kooijWhatDeterminesDrop2018} for flat-fan sprays, using the unstable wave-mediated sheet disintegration mechanism, and for the fragmenting turbulent jet by \citet{villermauxFragmentationCohesion2020} by balancing recombination and separation times of constituent blobs.

A temporal evolution of average droplet diameters is presented in figure~\ref{fig:scaling}b with definition $\langle d\rangle_t(\tau_{b,i}) \equiv \langle d\rangle(\tau_b\in Ti)$, which represents the average diameter at $\tau_{b,i}$, measured using the observed droplets in the corresponding time sub-period $Ti$. Here, the global or overall average $\langle d\rangle$ for the entire breakup period is used for normalization, revealing a similar trend for different $We$ in figure~\ref{fig:scaling}b. The smaller-sized droplets appear first (also see figure~\ref{fig:pdf-map}b), possibly since the unsteady impulsive airflow has a higher effective $We$ during early interaction. The number of daughter droplets $N$ can also be deduced by extending the analysis
\begin{equation}
    N \sim (r_0^3 /r_f^3 ) \sim We
\end{equation}
We consider the number of droplets produced per unit measurement volume, i.e., the number concentration $N_v=N/V_{\rm meas}$ for unbiased estimates of droplet count as discussed earlier \citep{sharmaDepthDefocusTechnique2023a, raoDepthDefocusTechnique2024a}. Temporal evolution of the normalized version $N_v/We$ is plotted in figure~\ref{fig:scaling}c, illustrating the validity of aforementioned scales (represented by matching peaks). A similar trend is observed across different $We$, with a large number of droplets generated in the early stages. Culminating both the ideas associated with the observed temporal evolution of $\langle d\rangle_t$ and $N_v$, we deduce that the droplets with smaller diameters appear first in large quantities. Since we observe an insignificant variation in the temporal evolution of $\langle d\rangle_t$ across all $We$ (see inset of figure~\ref{fig:scaling}b), it suggests that the overall size distributions are predominantly determined by the scales associated with the maxima of $N_v$ (see inset of figure~\ref{fig:scaling}c). The droplet production rate can be defined as $\nu=dN/dt$, which will be used next to derive a Time Integrated PDF.

The proposed scaling laws deduce averages without employing any specific mechanism of disintegration. A similar approach was recently proposed by \citet{villermauxFragmentationPrinciplesMechanisms2025a} for fragmentation with random processes, where generalized conservation laws alongside the maximal randomness principle were used to deduce a power-law for the fragment size distribution of the form $p(d)\sim d^{-\beta}$, where $\beta$ is a function of the dimensionality of the object. For the present case of aerobreakup, the fragmentation is three-dimensional, for which $\beta=3.5$. The corresponding power-law is illustrated in figure~\ref{fig:pdf}b, which is in good agreement with the experimental data. 

\subsection{Time Integrated Distribution}
As indicated through the plots in figure~\ref{fig:scaling}, the average droplet diameter and the production rates of the droplets show a transient behavior. Since the overall distribution is a cumulative representation of all these droplets produced over the breakup time period, an integrated approach is required. For a time period from $t$ to $t+\delta t$, the incremental number of droplets produced is given by
\begin{equation}
    \delta N \left(t\right)\approx\nu\left(t\right)\cdot \delta t
\end{equation}
For this period, the droplet size distribution is given using the aforementioned equations~\ref{eq:compound_gamma} and \ref{eq:normalized_pdf}, where $\left\langle d\right\rangle (t)$ is a function of time. The cumulative distribution of droplets or the Time Integrated Distribution (TID) $p^T\left(d\right)$ using the number of droplets generated in each time period as the weight function will be given as
\begin{equation} \label{eq:TID}
    p^T\left(d\right)=\frac{\int_{t_1}^{t_2}{\frac{1}{\left\langle d\right\rangle}\bar{p}\left(\frac{d}{\left\langle d\right\rangle}\right)} \nu dt}{\int_{t_1}^{t_2}\nu dt}
\end{equation}
where $\bar{p}\left(d/\left\langle d\right\rangle\right)$ is the universal compound gamma distribution deduced earlier and $t_1$ and $t_2$ correspond to the start and end of the breakup, respectively. In the normalized form
\begin{equation}
    {\bar{p}}^T\left(\frac{d}{\left\langle d\right\rangle^T}\right)=\left\langle d\right\rangle^T\cdot p^T\left(d\right)
\end{equation}
where $\left\langle d\right\rangle^T$ is the overall average diameter for TID given by
\begin{equation}
    \left\langle d\right\rangle^T=\int{d'\cdot p^T\left(d'\right)}\ d d'
\end{equation}
To estimate ${\bar{p}}^T$, the scaling laws are required for $We(t)$, $\langle d\rangle (t)$ and $\nu(t)$. Using the earlier established scaling laws and measured power-law decay of the airflow $u_a \propto t^\kappa$ discussed in the Appendix~\ref{apx:airflow}, we get
$\left\langle d\right\rangle/d_0 \sim {We}^{-1/3} \propto t^{-2\kappa/3}$ and $N \sim We\propto t^{2\kappa}$, where $\kappa=-1.75$. Hence, the droplet production rate is given by
$\nu=dN/dt\sim\ t^{2\kappa-1}$. Substituting this in equation \ref{eq:TID}, a TID is derived and is illustrated in figure~\ref{fig:pdf}. This distribution illustrates a good match at the tail in figure~\ref{fig:pdf}a, as also supported by goodness of fit estimates discussed in the Appendix~\ref{apx:dist}, indicating the validity of the proposed model. This approach can even be extended to systems where $\bar{p}$ is not steady but can be determined as a function of time.

\subsection{On generalized integrated approaches}
\label{sec:generalized}

Considering aerobreakup at lower Weber numbers, there is a tendency for the liquid to preferentially distribute into specific modes, such as the rim and bag films in bag breakup, leading to multimodal breakup behavior. \citet{jackiwAerodynamicDropletBreakup2021b, jackiwPredictionDropletSize2022b, jackiwAerodynamicDropletAtomization2023c} deduced size distribution through a cumulative approach, accounting for droplets produced by these different mechanisms in both bag breakup and sprays generated from twin-fluid nozzles. The current scenario of catastrophic breakup is characterized by chaotic emergence and breakup of undulations. While sub-secondary breakup processes can be associated with distinct modes, the overall droplet size distribution tends to be unimodal due to the inherent randomness of these events.

Continuing the discussion corresponding to figure~\ref{fig:stripping}, the recurrent breakup of the undulation results in the formation of daughter droplets. For instance, the $i$-th undulation generates a droplet cluster with droplet count $N_i$ and size distribution $p^i(d)$. Then, the overall droplet size distribution $p^I\left(d\right)$ can be deduced by using a simple integrated approach
\begin{equation}
    p^I\left(d\right)=\frac{\sum_{i}N_i \cdot p^i(d)}{\sum_{i}N_i}
\end{equation}
 which is similar to equation~\ref{eq:TID}. Although simplistic, such integrated approaches present an interesting perspective. If $p^i(d)$ and $N_i$ can be presented as a function of undulation size $\xi_i$ and local effective Weber number $We_i'$, then all we need is the statistics of $\xi_i$ and $We_i'$ to predict the overall distribution, where the occurrence frequencies of both enable the integration in space (representing the evolving droplet interface) and time (total breakup period). This approach is analogous to a sea spray function \citep{deikeMassTransferOcean2022, mazzatentaLinkingEmittedDrops2025}, where the wave-breaking probabilities, associated mode of droplet generation, and corresponding size distribution are integrated, along with other factors originating from environmental conditions. This function predicts air-sea exchange in terms of droplet and bubble formation, and the phenomenon of ocean waves shares the same multiscale nature of deformation, forming a deformation cascade. Hence, it is fascinating that the surface of a droplet facing catastrophic breakups and breaking waves at the sea surface shares the same fate philosophically. Interestingly, the sub-scale breakup topologies for high-speed wind-ocean interaction are also similar to the sub-secondary atomization modes discussed earlier \citep{veronSeaSpraySpume2012, troitskayaBagbreakupFragmentationDominant2017}.

To deduce a similar integrated function for the catastrophic breakup event, further information is necessary on the statistics of the turbulent interface dynamics involving the afore-mentioned undulation breakups, corresponding breakup modes, and daughter droplet sizes. Achieving this is extremely challenging both experimentally and numerically, owing to the large range of scales in the phenomenon and the need to resolve the interface at such small spatiotemporal scales.

\section{On self-similar intermediate interfacial topology}
\label{sec:selfsimilar}
\textcolor{black}{In the present work, self-similarity across scales was demonstrated qualitatively through the observed undulation breakup topology, an approach commonly adopted for identifying breakup regimes in both primary and secondary atomization. However, other manifestations of self-similarity were established quantitatively through the statistics of ligament corrugations and daughter droplet size distributions. The convergence of statistical metrics describing ligament corrugations provides quantitative evidence that the terminal liquid features preceding fragmentation possess statistically invariant geometrical characteristics across scales. Likewise, the collapse of the daughter droplet size distributions onto a common normalized PDF demonstrates the existence of a universal statistical description, where the actual distribution is recovered through scaling by the characteristic (mean) droplet diameter, analogous to similarity solutions encountered in classical fluid mechanics (e.g., boundary layers, pinch-offs, etc.). In this context, self-similarity of the ligament corrugations implies that the perturbations are random but statistically indistinguishable, whereas for the droplet size distributions, it implies that a single universal PDF describes fragmentation over a range of conditions after appropriate scaling. These observations collectively suggest the existence of a self-similar intermediate fragmentation state, characterized by statistically invariant interfacial features spanning multiple scales.}

\textcolor{black}{However, topological or geometric descriptors might be more effective indicators of the self-similar nature of disintegrating liquid interfaces. Various quantitative measures, often statistical in nature, have been employed historically to assess the extent of self-similarity. The elevation spectra of a liquid interface are a common place to study wave turbulence, and universal spectral power laws are derived and observed, representing a self-similar cascade across scales \citep{nazarenkoWaveTurbulenceWater2016, newellWaveTurbulence2011, falconExperimentsSurfaceGravity2022a}. Various regimes have been identified in wave turbulence within the weakly nonlinear regime and have also been employed extensively for ocean waves. However, an interface undergoing a catastrophic breakup enters a nonlinear regime well beyond the current models. Interface curvature is another tool; however, it is well-defined only for smooth manifolds, whereas events such as pinch-off introduce unavoidable singularities in atomization processes. Curvature statistics can be described for intermediate interfacial states and are closely linked to the droplet size distribution in the final state \citep{canuWhereDoesDroplet2018, vegadCurvaturebasedAnalysisComplex2026}. Consequently, they can serve as a valuable indicator of the evolution of the liquid body, transitioning from an initial, smooth state characterized by an approximately unimodal curvature distribution for the parent droplet, following the atomization process that broadens the distribution, to a stable final distribution that directly correlates with the resulting droplet sizes. The curvature metrics involve second-order derivatives and thus entail significant error propagation, necessitating very high measurement resolution. The self-similar nature across scales can also be captured by estimating the fractal dimension of the projected interface \citep{sreenivasanFractalsMultifractalsFluid1991,shavitFRACTALDIMENSIONSLIQUID1995, groutFractalAnalysisAtomizing2007}. The extent of self-replicating statistical behavior is then indicated by a fractal dimension between $1-2$ for a two-dimensional projection, as has been employed to study the turbulent interface of gaseous jets and liquid sprays.}

\textcolor{black}{Estimating these topological descriptors and spectral signatures requires high measurement resolution, which is severely inaccessible for the rapidly evolving finite boundary of a tiny droplet considered in the present study. The authors, therefore, are studying a parallel system involving the shock-induced disintegration of a liquid jet, which provides a simplified initial interface and larger spatiotemporal measurement window with reasonable resolution. The self-similar cascade is assessed through a universal power-law in the spectral signature, curvature PDFs, and fractal dimensions.}

\section{Conclusion}
\label{sec:conclusion}
In the present study, we have a closer look at the intermediate sub-scale processes that occur during the unsteady aerobreakup of a droplet in the range $We \approx 900-4000$ using a shock tube system. The subsequent multiscale interfacial deformations, originating from various unstable wave mechanisms (KHI, RTI, and RPI), form a deformation cascade. The undulations disintegrate through sub-secondary breakup processes, primarily conforming to a ligament or bag mode. The daughter droplet size distributions were estimated using a DFD technique, in which the normalized distributions at different $We$ and time periods were found to follow a universal distribution \textcolor{black}{within the investigated range}. The associated compound gamma distribution and its parameters align well with the coefficient determined from the terminal ligaments. This corresponds to the limit associated with the most extreme corrugations physically possible. This illustrates self-similar behavior pertaining to the transient size distributions at different $We$. \textcolor{black}{It is hypothesized that the consistent evolution of topological features and breakup morphology at different scales is bridged by effective dimensionless numbers such as $We'$ and $Oh'$, however further investigation is required in this direction.} Scaling laws for average droplet diameter and number of fragments were deduced based on a chaotic dispersal of disruptive energy. Enforcing these laws, the model for universal distribution was improved by adopting a time integrated approach, which predicted a larger tail and aligned well with the experimental data. 

To summarize, the self-similarity covers the following mechanisms: (a) \textcolor{black}{Similar} breakup topology at multiple scales, i.e., secondary atomization at droplet level and the sub-secondary breakup at undulation level, \textcolor{black}{explained using} effective Weber number  (b) The terminal ligaments are topologically similar since their corrugation are saturated to a \textcolor{black}{common limit with random protoblob sizes} (c) A universal daughter droplet size distribution for different aerodynamic strength $(We)$ and time instance.

%The quantitative reinforcement of self similar aspects needs further investigation. The multiscale nature of consistent breakup morphology, as observed in secondary atomization, sub-secondary atomization, and even the breakup of protrusions over larger entities such as liquid jets and ocean surfaces, can 0. 

% \backsection[Supplementary data]{\label{SupMat}Supplementary material and movies are available at \\https://doi.org/10.1017/jfm.2019...}

\backsection[Acknowledgements]{S.J.R. would like to thank the Prime Minister Research Fellowship (PMRF) for the financial support. S.B. would like to acknowledge the support from the Indian National Academy of Engineering (INAE) Chair professorship.}

% \backsection[Funding]{S.J.R. acknowledges funding through the Prime Minister’s Research Fellowship (PMRF).}

\backsection[Declaration of interests]{The authors report no conflict of interest.}

% \backsection[Data availability statement]{The data that support the findings of this study are openly available in [repository name] at http://doi.org/[doi], reference number [reference number]. See JFM's \href{https://www.cambridge.org/core/journals/journal-of-fluid-mechanics/information/journal-policies/research-transparency}{research transparency policy} for more information}

\backsection[Author ORCIDs]{S. J. Rao, https://orcid.org/0000-0001-6539-5814; S. Basu, https://orcid.org/0000-0002-9652-9966}

% \backsection[Author contributions]{Authors may include details of the contributions made by each author to the manuscript'}

\appendix

\section{Sub-secondary breakup events}\label{apx:subsec}

\begin{figure} 
  \centerline{\includegraphics[width=0.85\linewidth]{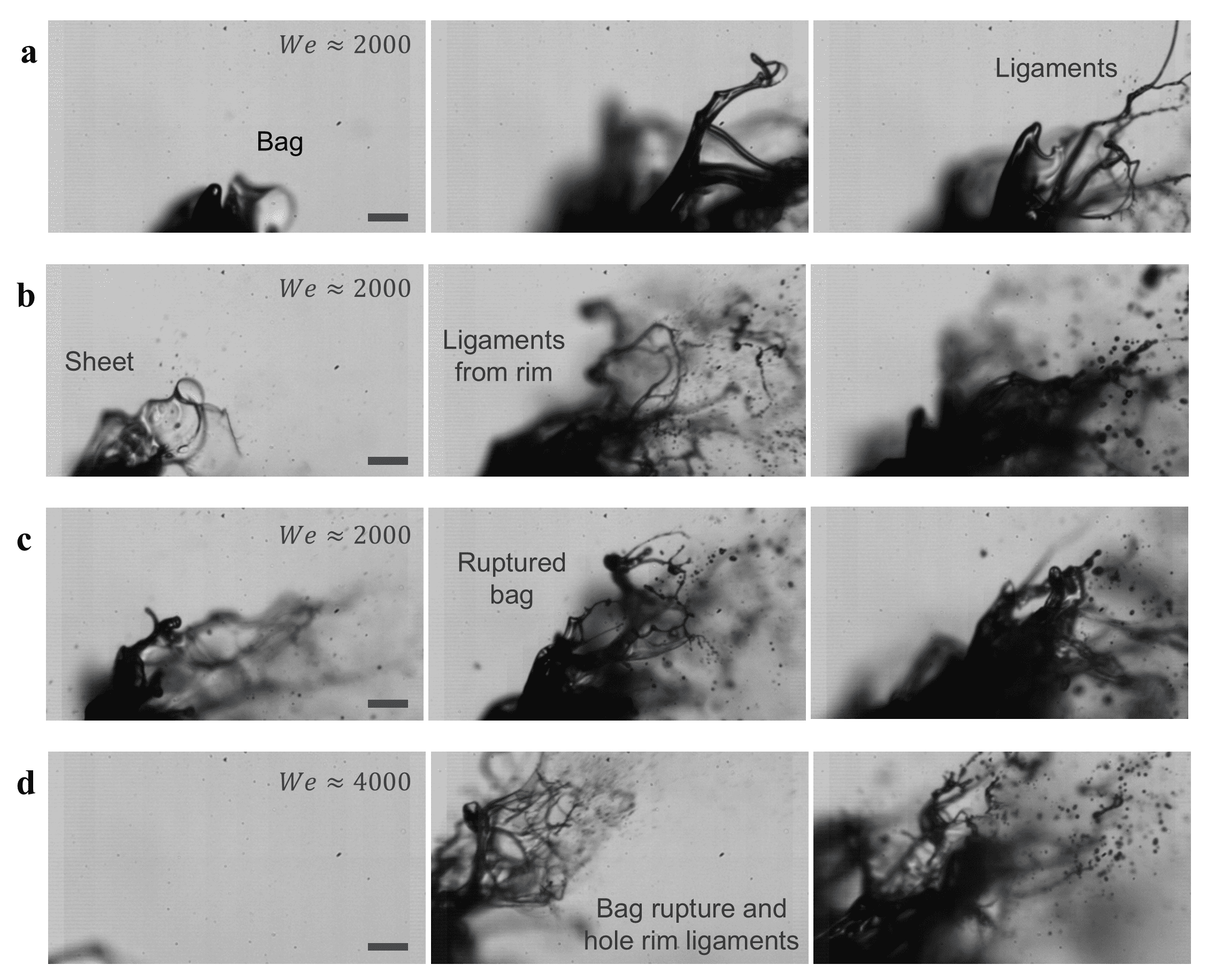}}
  \caption{Sub-secondary breakup processes showcasing the bag mode. The scale bar represents $100\mu m$ and consecutive frames are separated by an interval of $\Delta t=13.33\mu s$.}
\label{fig:z-sub-sec}
\end{figure}

The sub-secondary breakup events observed in the present range of global Weber numbers skew more towards the appearance of bag modes compared to the ligament mode. More of these events illustrating bag mode are shown in figure~\ref{fig:z-sub-sec}. \textcolor{black}{The estimates from the scaling arguments for local effective $We^\prime$ also suggest the sub-secondary breakup modes to reside in the bag mode. Across all the observations made, only $\sim 10\%$ of the events followed a ligament mode.}

\section{Unsteady airflow characteristics}\label{apx:airflow}

\begin{figure} 
  \centerline{\includegraphics[width=0.75\linewidth]{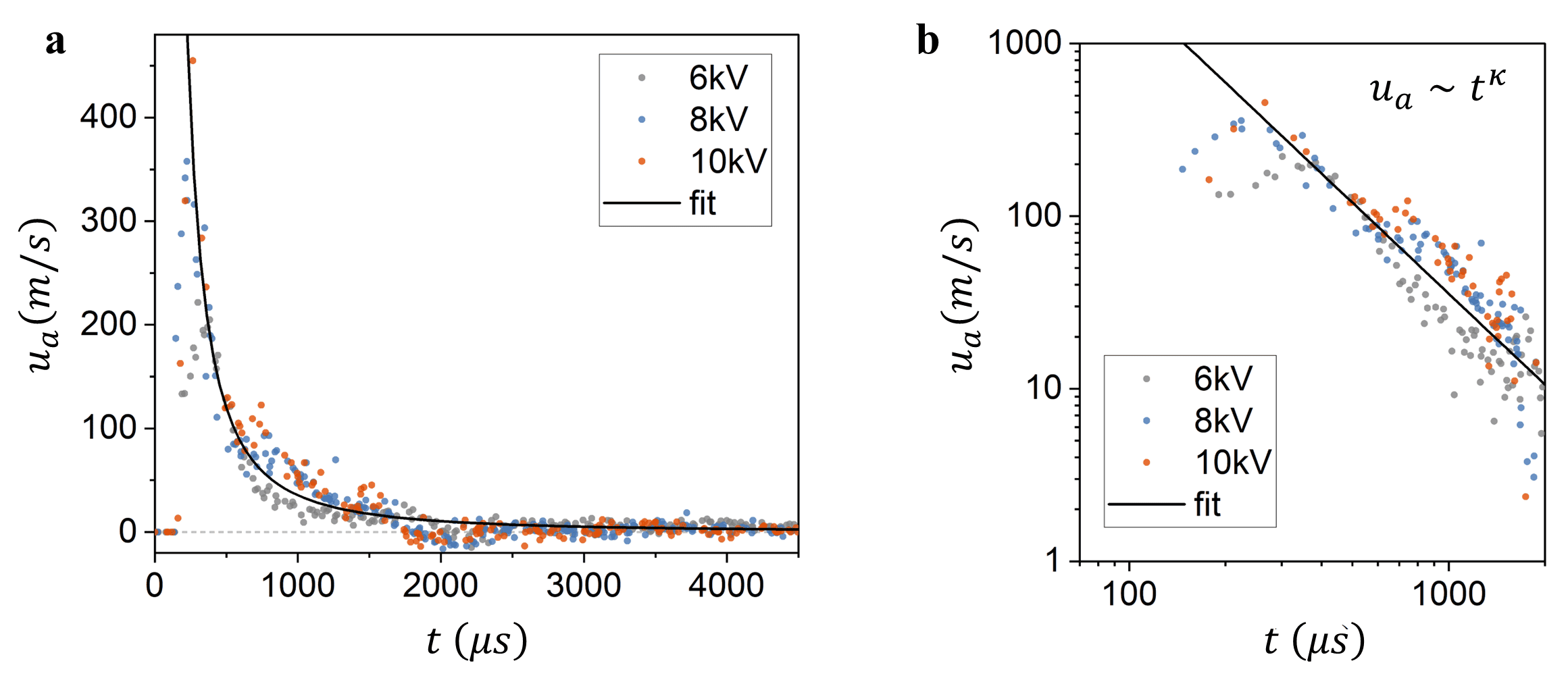}}
  \caption{(a) Axial airflow velocity outside the shock tube (b) Power-law decay of the axial airflow velocity outside the shock tube with $\kappa = -1.75$}
\label{fig:z-airflow}
\end{figure}

 The imposed disruptive energy originates from a decaying flow field in the present system, which deploys a wire explosion-based shock tube. The explosion generates a blast wave that interacts with the tube opening, resulting in a complex flow field exhibiting a characteristic decay. This involves compressible structures such as expansion waves, secondary shock and vortices. The axial air flow velocities were measured by \citet{raoBlastWaveInduced2025a} at the location of the droplet interaction using particle image velocimetry (PIV). The obtained flow field for different charging voltages is illustrated in figure~\ref{fig:z-airflow}a. An interesting observation is that the decay part of the flow in all cases overlaps, and only the impulsive jump in velocity corresponding to the first instance increases with increasing charging voltage. When plotted on a log scale, we see a power-law decay of the following form for all the cases
 \begin{equation}
     u_a=\hat{u}t^\kappa
 \end{equation}
For the present shock tube setup, the PIV measurements give $\kappa=-1.75$ as illustrated in figure~\ref{fig:z-airflow}b. From the best fit, we also have $\hat{u}=0.0002$ when time $t$ is expressed in seconds. Since $We\propto u_a^2$ we have $We\propto t^{2\kappa}$.

\begin{figure} 
  \centerline{\includegraphics[width=0.8\linewidth]{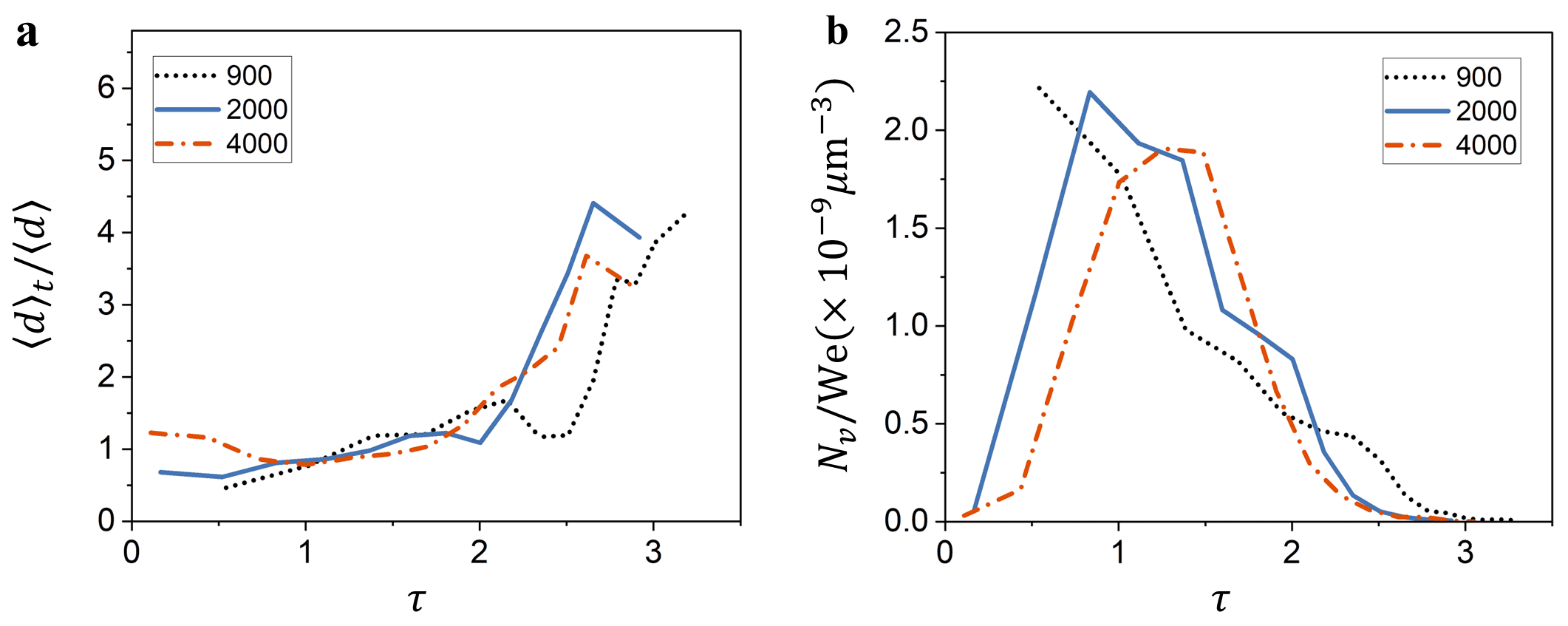}}
  \caption{Transient evolution of (a) average diameter of daughter droplets $\left\langle d\right\rangle_t/\left\langle d\right\rangle$ with $We$ (b) normalized droplet frequency $N_v/We$.  The time axis (with respect to shock-droplet interaction) is normalized using inertial time scales.}
\label{fig:z-timeshift}
\end{figure}

 To determine the instantaneous Weber number for each breakup period rather than a global value, we need to trace the droplets backward in time to the instant when the actual fragmentation occurred. For a time period $T_i$, let the first instant in the measurement region be $t_i^\ast$. Corresponding to this, we need to estimate the time $t_i$ at which the daughter droplet was actually produced via fragmentation. This lag exists because it requires time for the droplet to travel to the observation window. For simplification, let's assume the fragmentation is instantaneous, the air velocity is unsteady but uniform, and there is no-slip between the daughter droplets and the air. Then we have
\begin{equation}
    \frac{dx}{dt}=u_a=\hat{u}t^\kappa
\end{equation}
where $x$ describes the position of the daughter droplets and $x=0$ corresponds to the original location of the parent droplet, and $x=x_m=30mm$ represents the measurement zone positioned downstream. Integrating the equation and rearranging yields
 \begin{equation}
     t_i=\left\{{t_i^\ast}^{\kappa+1}-\frac{x_m\left(\kappa+1\right)}{\hat{u}}\right\}^\frac{1}{\kappa+1}
 \end{equation}
Using $u_i=\hat{u}t_i^\kappa$, we can estimate the corresponding air flow velocity and hence the instantaneous Weber number $ We_i \propto u_i^2$ for each time period, with the air density set to the unaffected ambient conditions \citep{raoBlastWaveInduced2025a}, and the same was used to illustrate the scaling law in figure~\ref{fig:scaling}a. This effect can also be accounted for the observed trends in figure~\ref{fig:scaling}b-c, where the time is shifted appropriately here and then normalized with the inertial time scale, as presented in figure~\ref{fig:z-timeshift}. Here, one observes a slightly better overlap between the normalized trends of $\langle d\rangle_t$ and $N_v$ for different flow strengths. Also, the breakup is found to be concluded within the period $\tau \approx 3$ for all the cases.

\section{Droplet size distribution fits}\label{apx:dist}
\textcolor{black}{This section includes the comparison of various PDFs commonly used in the atomization literature with the experimental data (normalized PDFs), such as lognormal (LN), root normal (RN) and gamma (G) distributions, alongside the theoretically derived compound gamma (CG) and Time Integrated Distribution (TID) in the present study. The equations of the PDFs are presented below:
\begin{equation} \label{eq:lognormal}
    \bar{p}_{ln}(\bar{d})=\frac{1}{\bar{d}\ \sqrt{2\pi \sigma_{ln}^2}}\exp{\left(- \frac{(\ln{\bar{d}}\ -\mu_{ln})^2}{2\sigma_{ln}^2 }\right)}
\end{equation}
\begin{equation} \label{eq:rootnormal}
    \bar{p}_{rn}(\bar{d})=\frac{1}{2\sqrt{\bar{d}\ }\ \sqrt{2\pi \sigma_{rn}^2}}\exp{\left(- \frac{({\sqrt{\bar{d}\ }}\ -\mu_{rn})^2}{2\sigma_{rn}^2 }\right)}
\end{equation}
\begin{equation} \label{eq:gamma}
    \bar{p}_{g}(\bar{d})=\frac{\bar{d}^{k-1}}{\Gamma(k)\ \theta^k}  \exp{\left(- \frac{\bar{d}}{\theta }\right)}
\end{equation}
\begin{figure} 
  \centerline{\includegraphics[width=1\linewidth]{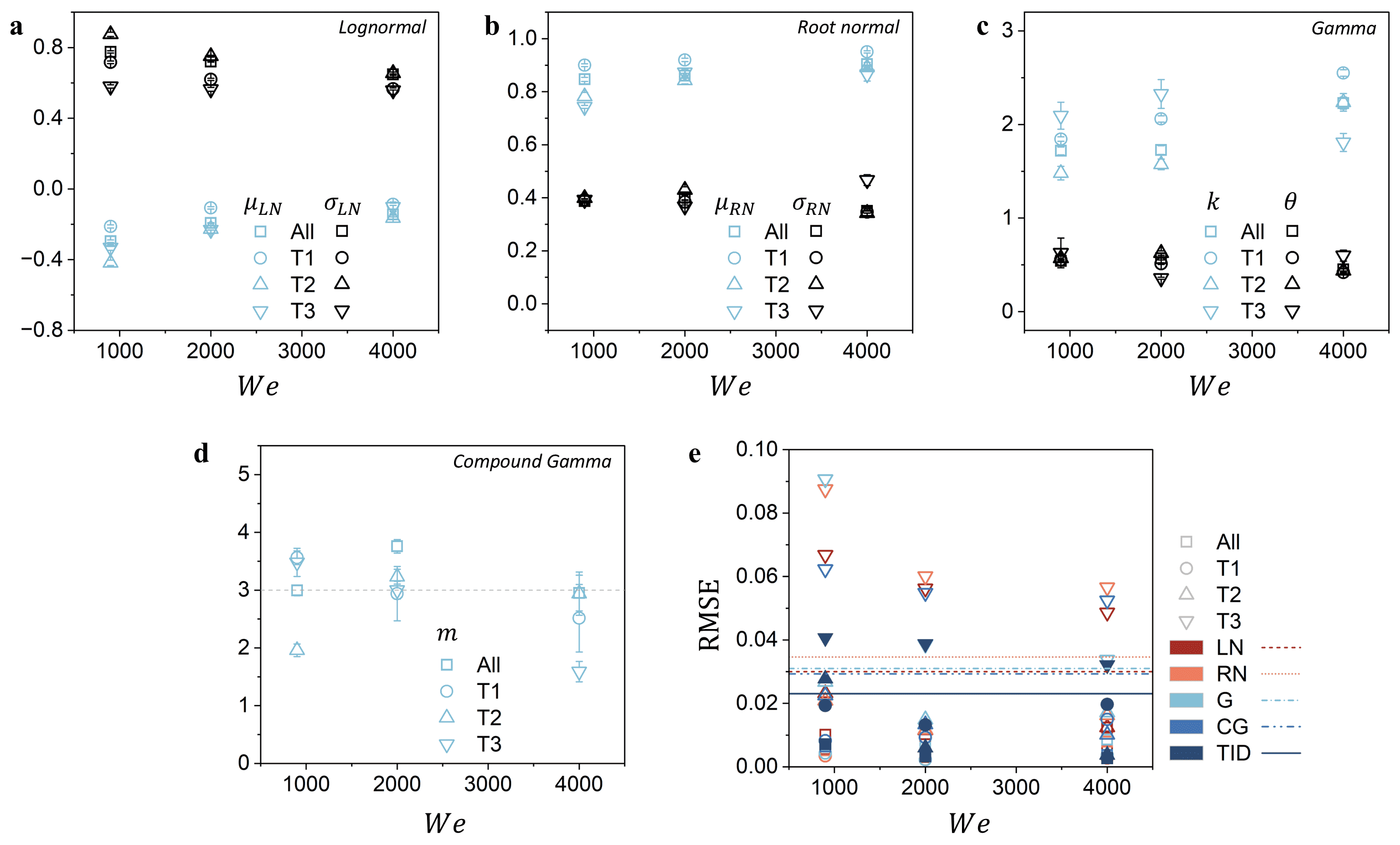}}
  \caption{Fit parameters with confidence intervals represented as error bars for (a) lognormal (LN), (b) root normal (RN), (c) gamma (G), and (d) compound gamma (CG) distributions. (e) Goodness-of-fit measures for the fits with respect to experimental distributions represented through RMSE, including the Time Integrated Distribution (TID).} 
\label{fig:distfit}
\end{figure}
where $\bar{d}$ is the normalized diameter. $\bar{p}_{ln}$, $\bar{p}_{rn}$, and $\bar{p}_{g}$ are the PDFs for LN, RN, and G distributions described by parameters $\{ \mu_{ln}, \sigma_{ln} \}$, $\{ \mu_{rn}, \sigma_{rn} \}$ and $\{ k, \theta \}$, respectively. The definitions of compound gamma and TID are given in equations~\ref{eq:compound_gamma} and \ref{eq:TID}, respectively. The fit parameters and corresponding confidence intervals (represented as error bars), and goodness-of-fit represented by Root Mean Square Error (RMSE) for these candidates at different $We$ and time periods are presented in figure~\ref{fig:distfit}. It also includes RMSE for the theoretically deduced TID. The fit parameters for most of these candidates are found to be stable within reasonable bounds for different flow conditions and time periods, as shown in figure~\ref{fig:distfit}a-d, supporting the universality of the PDFs within the investigated range. The RMSE values for all these PDFs are shown in figure~\ref{fig:distfit}e for convenient comparison. A deviation is uniformly observed for T3, and, as discussed earlier, abnormal behavior is expected at the end of the breakup period, as the flow decays. To compare the different fit types, the Root Mean Square (RMS) of all the RMSE values (T1-T3) for each case is shown as horizontal lines in figure~\ref{fig:distfit}e. TID is slightly superior to the rest, with LN, G, RN, and CG distributions falling short, failing to capture the long tails of the PDFs. The LN distribution has been widely accepted by the community as a suitable model for representing these distributions and also shows a better alignment at the tails relative to the CG distribution in figure~\ref{fig:pdf}. Nonetheless, the CG distribution provides a conceptual link between ligament topology and distribution, which is crucial for connecting intermediate interfacial states to final statistics. Although efforts have been initiated to establish a comparable connection between the LN distribution and corrugated ligaments \citep{palStatisticsDropsGenerated2024a}, further development is required to achieve a more comprehensive integration beyond its current statistical emphasis.}

\bibliographystyle{jfm}
\bibliography{jfm}

@article{cokeletBreakingWaves1977,
  title = {Breaking Waves},
  author = {Cokelet, E. D.},
  year = 1977,
  month = jun,
  journal = {Nature},
  volume = {267},
  number = {5614},
  pages = {769--774},
  issn = {1476-4687},
  doi = {10.1038/267769a0},
}

@article{young2025droplet,
  title={Droplet breakup by multimodal nonlinear Rayleigh Taylor instability},
  author={Young, Calvin J and Cook, Andrew W and McFarland, Jacob A},
  journal={International Journal of Multiphase Flow},
  pages={105490},
  year={2025},
  publisher={Elsevier}
}

@article{clauser1954turbulent,
  title={Turbulent boundary layers in adverse pressure gradients},
  author={Clauser, Francis H},
  journal={Journal of the Aeronautical Sciences},
  volume={21},
  number={2},
  pages={91--108},
  year={1954}
}

@article{canuWhereDoesDroplet2018,
	title = {Where does the droplet size distribution come from?},
	volume = {107},
	issn = {0301-9322},
	doi = {10.1016/j.ijmultiphaseflow.2018.06.010},
	journal = {International Journal of Multiphase Flow},
	author = {Canu, Romain and Puggelli, Stefano and Essadki, Mohamed and Duret, Benjamin and Menard, Thibaut and Massot, Marc and Reveillon, Julien and Demoulin, F. X.},
	month = oct,
	year = {2018},
	pages = {230--245},
}

@article{vegadCurvaturebasedAnalysisComplex2026,
	title = {Curvature-based analysis of complex morphologies at early stages of atomising spray},
	volume = {1027},
	issn = {0022-1120, 1469-7645},
	doi = {10.1017/jfm.2025.11110},
	journal = {Journal of Fluid Mechanics},
	author = {Vegad, Chetankumar S. and Ferrando, Diego and Carreres, Marcos and Idlahcen, Said and Duret, Benjamin and Reveillon, Julien and Demoulin, François-Xavier},
	month = jan,
	year = {2026},
	pages = {A43},
}

@article{falconExperimentsSurfaceGravity2022a,
	title = {Experiments in {Surface} {Gravity}–{Capillary} {Wave} {Turbulence}},
	volume = {54},
	issn = {0066-4189, 1545-4479},
	doi = {10.1146/annurev-fluid-021021-102043},
	number = {1},
	journal = {Annual Review of Fluid Mechanics},
	author = {Falcon, Eric and Mordant, Nicolas},
	month = jan,
	year = {2022},
	pages = {1--25},
}

@article{newellWaveTurbulence2011,
	title = {Wave {Turbulence}},
	volume = {43},
	issn = {0066-4189, 1545-4479},
	doi = {10.1146/annurev-fluid-122109-160807},
	number = {1},
	journal = {Annual Review of Fluid Mechanics},
	author = {Newell, Alan C. and Rumpf, Benno},
	month = jan,
	year = {2011},
	pages = {59--78},
}

@article{nazarenkoWaveTurbulenceWater2016,
	title = {Wave {Turbulence} on {Water} {Surface}},
	volume = {7},
	issn = {1947-5454, 1947-5462},
	doi = {10.1146/annurev-conmatphys-071715-102737},
	number = {1},
	journal = {Annual Review of Condensed Matter Physics},
	author = {Nazarenko, Sergey and Lukaschuk, Sergei},
	month = mar,
	year = {2016},
	pages = {61--88},
}

@article{groutFractalAnalysisAtomizing2007,
	title = {Fractal analysis of atomizing liquid flows},
	volume = {33},
	copyright = {https://www.elsevier.com/tdm/userlicense/1.0/},
	issn = {03019322},
	doi = {10.1016/j.ijmultiphaseflow.2007.02.009},
	number = {9},
	journal = {International Journal of Multiphase Flow},
	author = {Grout, Sébastien and Dumouchel, Christophe and Cousin, Jean and Nuglisch, Hans},
	month = sep,
	year = {2007},
	pages = {1023--1044},
}

@article{sreenivasanFractalsMultifractalsFluid1991,
	title = {Fractals and {Multifractals} in {Fluid} {Turbulence}},
	volume = {23},
	issn = {0066-4189, 1545-4479},
	doi = {10.1146/annurev.fl.23.010191.002543},
	number = {Volume 23, 1991},
	journal = {Annual Review of Fluid Mechanics},
	publisher = {Annual Reviews},
	author = {Sreenivasan, K. R.},
	month = jan,
	year = {1991},
	pages = {539--604},
}

@article{shavitFRACTALDIMENSIONSLIQUID1995,
	title = {Fractal dimensions of liquid jet interface under breakup},
	volume = {5},
	issn = {1044-5110},
	doi = {10.1615/AtomizSpr.v5.i6.10},
	number = {6},
	journal = {Atomization and Sprays},
	author = {Shavit, Uri and Chigier, Norman},
	year = {1995},
	pages = {525--543},
}

@article{alisedaXRayFlowVisualization2021,
  title = {X-{{Ray Flow Visualization}} in {{Multiphase Flows}}},
  author = {Aliseda, Alberto and Heindel, Theodore J.},
  year = 2021,
  journal = {Annual Review of Fluid Mechanics},
  volume = {53},
  pages = {543--567},
  issn = {0066-4189, 1545-4479},
  doi = {10.1146/annurev-fluid-010719-060201},
}

@article{tangBagFilmBreakup2023,
  title = {Bag Film Breakup of Droplets in Uniform Airflows},
  author = {Tang, K. and Adcock, T. A. A. and Mostert, W.},
  year = 2023,
  month = sep,
  journal = {Journal of Fluid Mechanics},
  volume = {970},
  pages = {A9},
  doi = {10.1017/jfm.2023.605},
}

@book{lefebvre2017atomization,
  title={Atomization and sprays},
  author={Lefebvre, Arthur H and McDonell, Vincent G},
  year={2017},
  publisher={CRC press}
}

@article{sharmaAdvancesDropletAerobreakup2022,
	title = {Advances in droplet aerobreakup},
	issn = {1951-6355, 1951-6401},
	url = {https://link.springer.com/10.1140/epjs/s11734-022-00653-z},
	doi = {10.1140/epjs/s11734-022-00653-z},
	journal = {The European Physical Journal Special Topics},
	author = {Sharma, Shubham and Chandra, Navin Kumar and Basu, Saptarshi and Kumar, Aloke},
	year = {2022},
}

@article{kooijWhatDeterminesDrop2018,
	title = {What {Determines} the {Drop} {Size} in {Sprays}?},
	volume = {8},
	doi = {10.1103/PhysRevX.8.031019},
	number = {3},
	urldate = {2024-02-08},
	journal = {Physical Review X},
	author = {Kooij, Stefan and Sijs, Rick and Denn, Morton M. and Villermaux, Emmanuel and Bonn, Daniel},
	year = {2018},
	pages = {031019},
}

@article{jackiwAerodynamicDropletAtomization2023c,
	title = {Aerodynamic droplet atomization model ({ADAM})},
	volume = {958},
	journal = {Journal of Fluid Mechanics},
	author = {Jackiw, Isaac M. and Ashgriz, Nasser},
	year = {2023},
	pages = {A2},
}

@article{dorschnerFormationRecurrentShedding2020a,
	title = {On the formation and recurrent shedding of ligaments in droplet aerobreakup},
	volume = {904},	
	journal = {Journal of Fluid Mechanics},
	author = {Dorschner, Benedikt and Biasiori-Poulanges, Luc and Schmidmayer, Kevin and El-Rabii, Hazem and Colonius, Tim},
	year = {2020},
	pages = {A20},
}

@article{guildenbecherSecondaryAtomization2009a,
	title = {Secondary atomization},
	volume = {46},
	doi = {10.1007/s00348-008-0593-2},
	number = {3},
	journal = {Experiments in Fluids},
	author = {Guildenbecher, D. R. and López-Rivera, C. and Sojka, P. E.},
	year = {2009},
	pages = {371--402},
}

@article{jackiwAerodynamicDropletBreakup2021b,
	title = {On aerodynamic droplet breakup},
	volume = {913},
	journal = {Journal of Fluid Mechanics},
	author = {Jackiw, Isaac M. and Ashgriz, Nasser},
	year = {2021},
	pages = {A33},
}

@article{jackiwPredictionDropletSize2022b,
	title = {Prediction of the droplet size distribution in aerodynamic droplet breakup},
	volume = {940},
	journal = {Journal of Fluid Mechanics},
	author = {Jackiw, Isaac M. and Ashgriz, Nasser},
	year = {2022},
	pages = {A17},
}

@article{jalaalTransientGrowthDroplet2014d,
	title = {Transient growth of droplet instabilities in a stream},
	volume = {26},
	doi = {10.1063/1.4851056},
	number = {1},
	journal = {Physics of Fluids},
	author = {Jalaal, M. and Mehravaran, K.},
	year = {2014},
	pages = {012101},
}

@article{marmottantSprayFormation2004b,
	title = {On spray formation},
	volume = {498},
	doi = {10.1017/S0022112003006529},
	journal = {Journal of Fluid Mechanics},
	author = {Marmottant, P. and Villermaux, E.},
	year = {2004},
	pages = {73--111},
}

@article{wangSimilarityPrimarySecondary2008a,
	title = {Similarity between the {Primary} and {Secondary} {Air}-{Assisted} {Liquid} {Jet} {Breakup} {Mechanisms}},
	volume = {100},
	doi = {10.1103/PhysRevLett.100.154502},
	number = {15},
	journal = {Physical Review Letters},
	author = {Wang, Yujie and Im, Kyoung-Su and Fezzaa, Kamel},
	year = {2008},
	pages = {154502},
}

@article{chandraShockinducedAerobreakupPolymeric2023a,
	title = {Shock-induced aerobreakup of a polymeric droplet},
	volume = {965},
	doi = {10.1017/jfm.2023.377},
	journal = {Journal of Fluid Mechanics},
	author = {Chandra, Navin Kumar and Sharma, Shubham and Basu, Saptarshi and Kumar, Aloke},
	month = jun,
	year = {2023},
	pages = {A1},
}

@article{sharmaDepthDefocusTechnique2023a,
	title = {Depth from defocus technique applied to unsteady shock-drop secondary atomization},
	volume = {64},
	doi = {10.1007/s00348-023-03588-w},
	number = {4},
	urldate = {2024-09-04},
	journal = {Experiments in Fluids},
	author = {Sharma, Shubham and Rao, Saini Jatin and Chandra, Navin Kumar and Kumar, Aloke and Basu, Saptarshi and Tropea, Cameron},
	year = {2023},
	pages = {65},
}

@article{sharmaShockInducedAerobreakup2021c,
	title = {Shock induced aerobreakup of a droplet},
	volume = {929},
	doi = {10.1017/jfm.2021.860},
	language = {en},
	urldate = {2024-09-04},
	journal = {Journal of Fluid Mechanics},
	author = {Sharma, Shubham and Pratap Singh, Awanish and Srinivas Rao, S. and Kumar, Aloke and Basu, Saptarshi},
	year = {2021},
	pages = {A27},
}

@article{sharmaShockinducedAtomisationLiquid2023a,
	title = {Shock-induced atomisation of a liquid metal droplet},
	volume = {972},
	doi = {10.1017/jfm.2023.705},
	journal = {Journal of Fluid Mechanics},
	author = {Sharma, Shubham and Chandra, Navin Kumar and Kumar, Aloke and Basu, Saptarshi},
	year = {2023},
	pages = {A7},
}

@article{raoDepthDefocusTechnique2024a,
	title = {Depth from defocus technique: a simple calibration-free approach for dispersion size measurement},
	volume = {65},
	doi = {10.1007/s00348-024-03792-2},
	number = {4},
	urldate = {2024-09-04},
	journal = {Experiments in Fluids},
	author = {Rao, Saini Jatin and Sharma, Shubham and Basu, Saptarshi and Tropea, Cameron},
	year = {2024},
	pages = {55},
}

@article{zandianUnderstandingLiquidjetAtomization2018,
	title = {Understanding liquid-jet atomization cascades via vortex dynamics},
	volume = {843},
	doi = {10.1017/jfm.2018.113},
	journal = {Journal of Fluid Mechanics},
	author = {Zandian, A. and Sirignano, W. A. and Hussain, F.},
	month = may,
	year = {2018},
	pages = {293--354},
}

@article{rimbertCrossoverRayleighTaylorInstability2011a,
	title = {Crossover between {Rayleigh}-{Taylor} instability and turbulent cascading atomization mechanism in the bag-breakup regime},
	volume = {84},
	doi = {10.1103/PhysRevE.84.016318},
	number = {1},
	journal = {Physical Review E},
	author = {Rimbert, Nicolas and Castanet, Guillaume},
	year = {2011},
	pages = {016318},
}

@article{villermauxLigamentMediatedSprayFormation2004,
	title = {Ligament-{Mediated} {Spray} {Formation}},
	volume = {92},
	doi = {10.1103/PhysRevLett.92.074501},
	number = {7},
	journal = {Physical Review Letters},
	author = {Villermaux, E. and Marmottant, Ph. and Duplat, J.},
	year = {2004},
	pages = {074501},
}

@article{zandianPlanarLiquidJet2017,
	title = {Planar liquid jet: {Early} deformation and atomization cascades},
	volume = {29},
	doi = {10.1063/1.4986790},
	number = {6},
	journal = {Physics of Fluids},
	author = {Zandian, A. and Sirignano, W. A. and Hussain, F.},
	year = {2017},
	pages = {062109},
}

@article{sultanovDropletSizeDistribution1990a,
	title = {Droplet size distribution in a percolation model for explosive liquid dispersal},
	volume = {31},
	doi = {10.1007/BF00852443},
	number = {5},
	journal = {Journal of Applied Mechanics and Technical Physics},
	author = {Sultanov, F. M. and Yarin, A. L.},
	year = {1990},
}

@article{zandianLengthscaleCascadeSpread2019,
	title = {Length-scale cascade and spread rate of atomizing planar liquid jets},
	volume = {113},
	doi = {10.1016/j.ijmultiphaseflow.2019.01.004},
	journal = {International Journal of Multiphase Flow},
	author = {Zandian, Arash and Sirignano, William A. and Hussain, Fazle},
	year = {2019},
	pages = {117--141},
}

@article{agbaglahBreakupThinLiquid2021,
	title = {Breakup of thin liquid sheets through hole–hole and hole–rim merging},
	volume = {911},
	doi = {10.1017/jfm.2020.1016},
	journal = {Journal of Fluid Mechanics},
	author = {Agbaglah, G. G.},
	year = {2021},
	pages = {A23},
}

@article{jeromeVorticesCatapultDroplets2013a,
	title = {Vortices catapult droplets in atomization},
	volume = {25},
	doi = {10.1063/1.4831796},
	number = {11},
	journal = {Physics of Fluids},
	author = {Jerome, J. John Soundar and Marty, Sylvain and Matas, Jean-Philippe and Zaleski, Stéphane and Hoepffner, Jérôme},
	year = {2013},
	pages = {112109},
}

@article{wangUniversalRimThickness2018a,
	title = {Universal {Rim} {Thickness} in {Unsteady} {Sheet} {Fragmentation}},
	volume = {120},
	url = {https://link.aps.org/doi/10.1103/PhysRevLett.120.204503},
	doi = {10.1103/PhysRevLett.120.204503},
	number = {20},
	journal = {Physical Review Letters},
	author = {Wang, Y. and Dandekar, R. and Bustos, N. and Poulain, S. and Bourouiba, L.},
	year = {2018},
	pages = {204503},
}

@article{hoepffnerSelfSimilarWaveProduced2011a,
	title = {Self-{Similar} {Wave} {Produced} by {Local} {Perturbation} of the {Kelvin}-{Helmholtz} {Shear}-{Layer} {Instability}},
	volume = {106},
	doi = {10.1103/PhysRevLett.106.104502},
	number = {10},
	journal = {Physical Review Letters},
	author = {Hoepffner, Jérôme and Blumenthal, Ralf and Zaleski, Stéphane},
	month = mar,
	year = {2011},
	pages = {104502},
}

@article{vadlamudiInsightsSpatiotemporalDynamics2024,
	title = {Insights into spatio-temporal dynamics during shock–droplet flame interaction},
	volume = {999},
	doi = {10.1017/jfm.2024.575},
	journal = {Journal of Fluid Mechanics},
	author = {Vadlamudi, Gautham and Aravind, Akhil and Rao, Saini Jatin and Basu, Saptarshi},
	month = nov,
	year = {2024},
	pages = {A22},
}

@article{keshavarzLigamentMediatedFragmentation2016,
	title = {Ligament {Mediated} {Fragmentation} of {Viscoelastic} {Liquids}},
	volume = {117},
	doi = {10.1103/PhysRevLett.117.154502},
	language = {en},
	number = {15},
	journal = {Physical Review Letters},
	author = {Keshavarz, Bavand and Houze, Eric C. and Moore, John R. and Koerner, Michael R. and McKinley, Gareth H.},
	month = oct,
	year = {2016},
	pages = {154502},
}

@article{veronSeaSpraySpume2012,
	title = {Sea spray spume droplet production in high wind speeds},
	volume = {39},
	doi = {10.1029/2012GL052603},
	number = {16},
	urldate = {2025-01-25},
	journal = {Geophysical Research Letters},
	author = {Veron, F. and Hopkins, C. and Harrison, E. L. and Mueller, J. A.},
	year = {2012},
}

@article{troitskayaBagbreakupFragmentationDominant2017,
	title = {Bag-breakup fragmentation as the dominant mechanism of sea-spray production in high winds},
	volume = {7},
	doi = {10.1038/s41598-017-01673-9},
	number = {1},
	journal = {Scientific Reports},
	author = {Troitskaya, Yu and Kandaurov, A. and Ermakova, O. and Kozlov, D. and Sergeev, D. and Zilitinkevich, S.},
	year = {2017},
	pages = {1614},
}

@article{chandraAerodynamicBagBreakup2024,
	title = {Aerodynamic bag breakup of a polymeric droplet},
	volume = {9},
	doi = {10.1103/PhysRevFluids.9.113303},
	number = {11},
	journal = {Physical Review Fluids},
	author = {Chandra, Navin Kumar and Sharma, Shubham and Basu, Saptarshi and Kumar, Aloke},
	month = nov,
	year = {2024},
	pages = {113303},
}

@article{chandraElasticityAffectsShockinduced2024,
	title = {Elasticity affects the shock-induced aerobreakup of a polymeric droplet},
	volume = {65},
	doi = {10.1007/s00348-024-03816-x},
	number = {5},
	journal = {Experiments in Fluids},
	author = {Chandra, Navin Kumar and Sharma, Shubham and Basu, Saptarshi and Kumar, Aloke},
	month = may,
	year = {2024},
	pages = {75},
}

@article{palStatisticsDropsGenerated2024a,
	title = {Statistics of drops generated from ensembles of randomly corrugated ligaments},
	volume = {36},
	doi = {10.1063/5.0221732},
	number = {11},
	journal = {Physics of Fluids},
	author = {Pal, Sagar and Pairetti, César and Crialesi-Esposito, Marco and Fuster, Daniel and Zaleski, Stéphane},
	month = nov,
	year = {2024},
	pages = {112116},
}

@article{neelFinesCollisionLiquid2020,
	title = {‘{Fines}’ from the collision of liquid rims},
	volume = {893},
	doi = {10.1017/jfm.2020.235},
	journal = {Journal of Fluid Mechanics},
	author = {Néel, B. and Lhuissier, H. and Villermaux, E.},
	month = jun,
	year = {2020},
	pages = {A16},
}

@article{theofanousAerobreakupNewtonianViscoelastic2011a,
	title = {Aerobreakup of {Newtonian} and {Viscoelastic} {Liquids}},
	volume = {43},
	doi = {10.1146/annurev-fluid-122109-160638},
	number = {1},
	urldate = {2024-09-04},
	journal = {Annual Review of Fluid Mechanics},
	author = {Theofanous, T.G.},
	month = jan,
	year = {2011},
	pages = {661--690},
}

@article{driessen2013stability,
  title={Stability of viscous long liquid filaments},
  author={Driessen, Theo and Jeurissen, Roger and Wijshoff, Herman and Toschi, Federico and Lohse, Detlef},
  journal={Physics of fluids},
  volume={25},
  number={6},
  year={2013},
  publisher={AIP Publishing}
}

@article{marmottant2004fragmentation,
  title={Fragmentation of stretched liquid ligaments},
  author={Marmottant, Philippe and Villermaux, Emmanuel},
  journal={Physics of fluids},
  volume={16},
  number={8},
  pages={2732--2741},
  year={2004},
  publisher={AIP Publishing}
}

@article{castrejon2012breakup,
  title={Breakup of liquid filaments},
  author={Castrej{\'o}n-Pita, Alfonso A and Castrej{\'o}n-Pita, JR and Hutchings, IM},
  journal={Physical review letters},
  volume={108},
  number={7},
  pages={074506},
  year={2012},
  publisher={APS}
}

@article{wang2021growth,
  title={Growth and breakup of ligaments in unsteady fragmentation},
  author={Wang, Y and Bourouiba, L},
  journal={Journal of Fluid Mechanics},
  volume={910},
  pages={A39},
  year={2021},
  publisher={Cambridge University Press}
}

@article{chenGalinstanLiquidMetal2018,
  title = {Galinstan Liquid Metal Breakup and Droplet Formation in a Shock-Induced Cross-Flow},
  author = {Chen, Yi and Wagner, Justin L. and Farias, Paul A. and DeMauro, Edward P. and Guildenbecher, Daniel R.},
  year = 2018,
  month = sep,
  journal = {International Journal of Multiphase Flow},
  volume = {106},
  pages = {147--163},
  issn = {03019322},
  doi = {10.1016/j.ijmultiphaseflow.2018.05.015},
}

@article{deikeMassTransferOcean2022,
  title = {Mass {{Transfer}} at the {{Ocean}}--{{Atmosphere Interface}}: {{The Role}} of {{Wave Breaking}}, {{Droplets}}, and {{Bubbles}}},
  shorttitle = {Mass {{Transfer}} at the {{Ocean}}--{{Atmosphere Interface}}},
  author = {Deike, Luc},
  year = 2022,
  month = jan,
  journal = {Annual Review of Fluid Mechanics},
  volume = {54},
  number = {Volume 54, 2022},
  pages = {191--224},
  publisher = {Annual Reviews},
  issn = {0066-4189, 1545-4479},
  doi = {10.1146/annurev-fluid-030121-014132},
}

@article{guildenbecherCharacterizationDropAerodynamic2017a,
  title = {Characterization of Drop Aerodynamic Fragmentation in the Bag and Sheet-Thinning Regimes by Crossed-Beam, Two-View, Digital in-Line Holography},
  author = {Guildenbecher, Daniel R. and Gao, Jian and Chen, Jun and Sojka, Paul E.},
  year = 2017,
  month = sep,
  journal = {International Journal of Multiphase Flow},
  volume = {94},
  pages = {107--122},
  issn = {03019322},
  doi = {10.1016/j.ijmultiphaseflow.2017.04.011},
}

@article{hsiangDropPropertiesSecondary1993,
  title = {Drop Properties after Secondary Breakup},
  author = {Hsiang, L. -P. and Faeth, G. M.},
  year = 1993,
  month = oct,
  journal = {International Journal of Multiphase Flow},
  volume = {19},
  number = {5},
  pages = {721--735},
  issn = {0301-9322},
  doi = {10.1016/0301-9322(93)90039-W},
}

@article{hsiangNearlimitDropDeformation1992b,
  title = {Near-Limit Drop Deformation and Secondary Breakup},
  author = {Hsiang, L. -P. and Faeth, G. M.},
  year = 1992,
  month = sep,
  journal = {International Journal of Multiphase Flow},
  volume = {18},
  number = {5},
  pages = {635--652},
  issn = {0301-9322},
  doi = {10.1016/0301-9322(92)90036-G},
}

@article{kamiyaStudyCharacteristicsFragment2022a,
  title = {Study on Characteristics of Fragment Size Distribution Generated via Droplet Breakup by High-Speed Gas Flow},
  author = {Kamiya, Tomohiro and Asahara, Makoto and Yada, Tokiha and Mizuno, Kyohei and Miyasaka, Takeshi},
  year = 2022,
  month = jan,
  journal = {Physics of Fluids},
  volume = {34},
  number = {1},
  pages = {012118},
  issn = {1070-6631},
  doi = {10.1063/5.0076448},
}

@article{kantBagmediatedFilmAtomization2023,
  title = {Bag-Mediated Film Atomization in a Cough Machine},
  author = {Kant, Pallav and Pairetti, Cesar and Saade, Youssef and Popinet, St{\'e}phane and Zaleski, St{\'e}phane and Lohse, Detlef},
  year = 2023,
  month = jul,
  journal = {Physical Review Fluids},
  volume = {8},
  number = {7},
  pages = {074802},
  publisher = {American Physical Society},
  doi = {10.1103/PhysRevFluids.8.074802},
}

@article{kuoMaximumEntropyFormalism2022a,
  title = {A Maximum Entropy Formalism Model for the Breakup of a Droplet},
  author = {Kuo, Chia-Wei and Trujillo, Mario F.},
  year = 2022,
  month = jan,
  journal = {Physics of Fluids},
  volume = {34},
  number = {1},
  pages = {013315},
  issn = {1070-6631},
  doi = {10.1063/5.0076910},
}

@article{mazzatentaLinkingEmittedDrops2025,
  title = {Linking Emitted Drops to Collective Bursting Bubbles across a Wide Range of Bubble Size Distributions},
  author = {Mazzatenta, Megan and Erinin, Martin A. and N{\'e}el, Baptiste and Deike, Luc},
  year = 2025,
  month = jul,
  journal = {Journal of Fluid Mechanics},
  volume = {1015},
  pages = {A8},
  issn = {0022-1120, 1469-7645},
  doi = {10.1017/jfm.2025.10273},
}

@article{nichollsAerodynamicShatteringLiquid1969,
  title = {Aerodynamic Shattering of Liquid Drops.},
  author = {Nicholls, J. A. and Ranger, A. A.},
  year = 1969,
  journal = {AIAA Journal},
  volume = {7},
  number = {2},
  pages = {285--290},
  publisher = {{American Institute of Aeronautics and Astronautics}},
  issn = {0001-1452},
  doi = {10.2514/3.5087},
}

@misc{raoBlastWaveInduced2025a,
  title = {Blast Wave Induced Unsteady Flow at the Shock Tube Opening},
  author = {Rao, Saini Jatin and Aravind, Akhil and Basu, Saptarshi},
  year = 2025,
  month = oct,
  number = {arXiv:2511.00247},
  eprint = {2511.00247},
  primaryclass = {physics},
  publisher = {arXiv},
  doi = {10.48550/arXiv.2511.00247},
}

@article{salauddinDetonationShockinducedBreakup2023,
  title = {Detonation and Shock-Induced Breakup Characteristics of {{RP-2}} Liquid Droplets},
  author = {Salauddin, S. and Morales, A. J. and Hytovick, R. and Burke, R. and Malik, V. and Patten, J. and Schroeder, S. and Ahmed, K. A.},
  year = 2023,
  month = apr,
  journal = {Shock Waves},
  volume = {33},
  number = {3},
  pages = {191--203},
  issn = {1432-2153},
  doi = {10.1007/s00193-023-01132-7},
}

@article{sellensSimplifiedPredictionDroplet1986,
  title = {A Simplified Prediction of Droplet Velocity Distributions in a Spray},
  author = {Sellens, R. W. and Brzustowski, T. A.},
  year = 1986,
  month = sep,
  journal = {Combustion and Flame},
  volume = {65},
  number = {3},
  pages = {273--279},
  issn = {0010-2180},
  doi = {10.1016/0010-2180(86)90041-6},
}

@article{villermauxDropFragmentationImpact2011,
  title = {Drop Fragmentation on Impact},
  author = {Villermaux, E. and Bossa, B.},
  year = 2011,
  month = feb,
  journal = {Journal of Fluid Mechanics},
  volume = {668},
  pages = {412--435},
  issn = {1469-7645, 0022-1120},
  doi = {10.1017/S002211201000474X},
}

@article{villermauxFragmentationCohesion2020,
  title = {Fragmentation versus {{Cohesion}}},
  author = {Villermaux, Emmanuel},
  year = 2020,
  month = sep,
  journal = {Journal of Fluid Mechanics},
  volume = {898},
  pages = {P1},
  issn = {0022-1120, 1469-7645},
  doi = {10.1017/jfm.2020.366},
}

@article{eggersPhysicsLiquidJets2008,
  title = {Physics of Liquid Jets},
  author = {Eggers, Jens and Villermaux, Emmanuel},
  year = 2008,
  month = mar,
  journal = {Reports on Progress in Physics},
  volume = {71},
  number = {3},
  pages = {036601},
  issn = {0034-4885, 1361-6633},
  doi = {10.1088/0034-4885/71/3/036601},
}

@article{simmonsCorrelationDropSizeDistributions1977a,
  title = {The {{Correlation}} of {{Drop-Size Distributions}} in {{Fuel Nozzle Sprays}}---{{Part I}}: {{The Drop-Size}}/{{Volume-Fraction Distribution}}},
  shorttitle = {The {{Correlation}} of {{Drop-Size Distributions}} in {{Fuel Nozzle Sprays}}---{{Part I}}},
  author = {Simmons, H. C.},
  year = 1977,
  month = jul,
  journal = {Journal of Engineering for Power},
  volume = {99},
  number = {3},
  pages = {309--314},
  issn = {0022-0825},
  doi = {10.1115/1.3446488},
}

@article{villermauxFragmentation2007,
  title = {Fragmentation},
  author = {Villermaux, E.},
  year = 2007,
  month = jan,
  journal = {Annual Review of Fluid Mechanics},
  volume = {39},
  number = {1},
  pages = {419--446},
  issn = {0066-4189, 1545-4479},
  doi = {10.1146/annurev.fluid.39.050905.110214},
}

@article{dworzanczykAerobreakupStagnationRegion2025,
  title = {On Aerobreakup in the Stagnation Region of High-{{Mach-number}} Flow over a Bluff Body},
  author = {Dworzanczyk, A. R. and {Viqueira-Moreira}, M. and Langhorn, J. D. and Libeau, M. A. and Brehm, C. and Parziale, N. J.},
  year = 2025,
  month = jan,
  journal = {Journal of Fluid Mechanics},
  volume = {1002},
  pages = {A1},
  issn = {0022-1120, 1469-7645},
  doi = {10.1017/jfm.2024.1092},
}

@article{villermauxFragmentationPrinciplesMechanisms2025a,
  title = {Fragmentation: {{Principles}} versus {{Mechanisms}}},
  shorttitle = {Fragmentation},
  author = {Villermaux, Emmanuel},
  year = 2025,
  month = nov,
  journal = {Physical Review Letters},
  volume = {135},
  number = {22},
  pages = {228201},
  publisher = {American Physical Society},
  doi = {10.1103/r7xz-5d9c},
}

@article{jonesFluidDynamicInduced2019,
  title = {Fluid Dynamic Induced Break-up during Volcanic Eruptions},
  author = {Jones, T. J. and Reynolds, C. D. and Boothroyd, S. C.},
  year = 2019,
  month = aug,
  journal = {Nature Communications},
  volume = {10},
  number = {1},
  pages = {3828},
  publisher = {Nature Publishing Group},
  issn = {2041-1723},
  doi = {10.1038/s41467-019-11750-4},
}

@article{raoSecondaryAtomizationDroplets2025c,
  title = {Secondary {{Atomization}} of {{Droplets}} at {{Extreme Conditions}}},
  author = {Rao, Saini Jatin and Basu, Saptarshi},
  year = 2026,
  volume = {58},
  pages = {83--110},
  month = aug,
  journal = {Annual Review of Fluid Mechanics},
  issn = {0066-4189, 1545-4479},
  doi = {10.1146/annurev-fluid-112823-115348},
}

@article{tropeaOpticalParticleCharacterization2011,
  title = {Optical {{Particle Characterization}} in {{Flows}}},
  author = {Tropea, Cameron},
  year = 2011,
  journal = {Annual Review of Fluid Mechanics},
  volume = {43},
  number = {1},
  pages = {399--426},
  doi = {10.1146/annurev-fluid-122109-160721},
}

@article{zhouSprayDropMeasurements2020,
  title = {Spray Drop Measurements Using Depth from Defocus},
  author = {Zhou, Wu and Tropea, Cameron and Chen, Benting and Zhang, Yukun and Luo, Xu and Cai, Xiaoshu},
  year = 2020,
  month = jul,
  journal = {Measurement Science and Technology},
  volume = {31},
  number = {7},
  pages = {075901},
  issn = {0957-0233, 1361-6501},
  doi = {10.1088/1361-6501/ab79c6},
}
%Use of the above commands will create a bibliography using the .bib file. Shown below is a bibliography built from individual items.

%\bibliographystyle{jfm}
%\bibliography{jfm2esam}

%% End of file `jfm2esam.bib'.

\end{document}